\documentclass[final,3p,times,twocolumn,sort&compress]{elsarticle}
\journal{Acta Materialia}

\usepackage[utf8]{inputenc}
\usepackage{hyperref}
\usepackage{graphicx}
\usepackage{subfig}
\usepackage{array}
\usepackage{amsmath,amssymb}
\usepackage{xspace}
\newcommand{\etal}{\textit{et al}.\@\xspace}

\newcommand{\cf}{\textit{cf}.\@\xspace}
\newcommand{\pwscf}{\textsc{Pwscf}\@\xspace}
\newcommand{\mcsqs}{\textsc{Mcsqs}\@\xspace}
\newcommand{\atat}{\textsc{Atat}\@\xspace}
\newcommand{\vasp}{\textsc{Vasp}\@\xspace}
\newcommand{\abinit}{\textit{ab initio}\@\xspace}

\newcommand{\rA}{\textrm{A}}
\newcommand{\rB}{\textrm{B}}
\newcommand{\rZr}{\textrm{Zr}}
\newcommand{\rNb}{\textrm{Nb}}

\newcommand{\rmix}{\textrm{mix}}
\newcommand{\rel}{\textrm{el}}
\newcommand{\rvib}{\textrm{vib}}
\newcommand{\md}{\mathrm{d}}

\begin{document}
%%%%%%%%%%%%%%%%%%%%%%%%%%%%%%%%%%%%%%%%%%%%%%%%%%%%%%%%%%%
%%%%%%%%%%%%%%%%%%%%%%%%%%%%%%%%%%%%%%%%%%%%%%%%%%%%%%%%%%%

\begin{frontmatter}

\title{Solubility in Zr-Nb alloys from first-principles}

\author{Maeva Cottura\fnref{UCB}}
\author{Emmanuel Clouet\corref{CA}}
\address{DEN-Service de Recherches de Métallurgie Physique, CEA, Université Paris-Saclay, F-91191 Gif-sur-Yvette, France}
\cortext[CA]{Corresponding author}
\ead{emmanuel.clouet@cea.fr}
\fntext[UCB]{Present address: Department of Materials Science and Engineering, University of California, Berkeley, CA 94720, USA}

\begin{abstract}
The thermodynamic properties of the Zr-Nb alloy are investigated at temperatures below 890\,K with \abinit calculations. 
The solution energies of the bcc Nb-rich and hcp Zr-rich solid solutions obtained within the framework of density functional theory are in good agreement with experimental data, although insufficient for a quantitative description of the miscibility gap, for which non configurational entropy has to be accounted for.
Whereas electronic free energies can be neglected, we show, using the harmonic approximation and the density functional perturbation theory, that both solution free energies are strongly modified by the contribution related to atomic vibrations.
Considering this vibrational free energy leads to a good description of the phase diagram.
\end{abstract}


\end{frontmatter}

\section{Introduction}

Zirconium alloys are structural materials with a wide range of technological applications,
in particular in the nuclear and chemical industries \cite{Northwood1985,Lemaignan2012}.
Among them, Zr-Nb alloys are model metallic systems which exhibit excellent mechanical properties,
associated with the fine microstructure resulting from different thermal annealing treatments, 
as well as superior corrosion resistance due to the formation of a passive oxide on the surface 
\cite{Allen2012,Motta2015,Matsukawa2017b}.
These made them, for instance, valuable materials for nuclear applications 
where they are used for the fuel assembly cladding in light water reactors
due to their excellent properties in service conditions.
Being biocompatible, they can also be employed in biomedical applications such as knee and hip replacement for orthopedic surgeries \cite{Zhou2012},
and their potential low magnetic susceptibility make them attractive for implants 
compatible with magnetic resonance imaging \cite{Nomura2009,Kondo2011}.
In addition, Zr-Nb multilayers have been investigated due to their superconducting properties
\cite{Lowe1984,Claeson1984},
or, more recently, for their possible superior mechanical performance and irradiation resistance
\cite{Thompson2003,Thompson2004,Frutos2015,Callisti2016,Carpenter2015,Ardeljan2016}.

Experimental studies of Zr-Nb alloys allow a clear identification of the microstructure
which corresponds at low temperature (below $\sim890$\,K) to an $\alpha$-Zr matrix,
with a hexagonal close-packed (hcp) crystallographic structure, containing
native $\beta$-niobium precipitates with a body centered cubic (bcc) crystallography. 
Because of the low mobility of Nb in the $\alpha$-Zr matrix, the alloy is usually not in its equilibrium state
and the matrix is supersaturated in Nb \cite{Toffolon2005,Onimus2012}.
Scarce particles of Laves phase containing niobium are also found in industrial alloys 
due to the presence of other alloying elements like Fe or Cr \cite{Toffolon2002,Barberis2004,Toffolon2005}. 
Under irradiation, one observes a partial amorphization and dissolution of these Laves phases,
which is common to all zirconium alloys,
and, more specifically for Zr-Nb alloys, the formation of very fine Nb-rich $\beta$ precipitates leading to a decrease of the Nb content in the $\alpha$ matrix 
\cite{Coleman1981,Cann1993,Perovic1993,Woo1994,Onimus2012,Bechade2013,Doriot2014,Menut2015}.
Since the mechanical properties of Zr-Nb alloys \cite{Yang2016,Matsukawa2016,Matsukawa2017a},
as well as their corrosion resistance \cite{Jeong2003a,Jeong2003b,Matsukawa2017b}
and their magnetic susceptibility \cite{Nomura2009,Kondo2011}, 
are strongly dependent on the precipitate microstructure and on the solute content of the $\alpha$ matrix, 
knowledge of the thermodynamic properties of the system as well as the resulting phase boundaries is of critical technological interest.

The experimental data for the Zr-Nb system display a non-symmetrical miscibility gap 
yielding a zero solubility limit in both rich regions of the phase diagram at 0\,K. 
Based on the Calphad assessments of Guillermet \cite{Guillermet1991} and of Lafaye \cite{Lafaye2017},
the Zr solubility in the $\beta$ Nb-rich phase is about 8\,at.\%
at the monotectoid temperature (890\,K),
whereas the Nb solubility in the $\alpha$ Zr-rich phase is much lower (0.8\,at\% Nb).
However, only partial agreement exists in the literature regarding a quantitative description of the Zr-Nb phase boundaries. 
Due to the low diffusivity of niobium in zirconium
and the even smaller one of zirconium in niobium,
it is difficult to reach equilibrium at low temperature.
Besides, the solubility limits are highly sensitive to the presence of impurities, 
in particular interstitial alloying elements such as oxygen
 \cite{Guillermet1991,Abriata1982,Okamoto1992,Toffolon2002,Kim2005},
but also substitutional impurities such as Fe or Cr \cite{Toffolon2002,Barberis2004,Toffolon2005,Kim2005}.
As a consequence, the measured values of the solubility limits at the monotectoid temperature 
spread from 6 to 15\,at.\% Zr in the $\beta$ phase
and from 0.5 to 1.5\,at.\% Nb in the $\alpha$ phase \cite{Abriata1982}.
The phase diagram of Guillermet \cite{Guillermet1991} is usually preferred
because of the thermodynamic self-consistency inherent to its Calphad approach.
The corresponding solubility limits have been assessed indirectly from selected pieces of experimental data
where exposure to impurities have been controlled and long annealing times applied \cite{Flewitt1972,VanEffenterre1972}.
The solubility limits in the temperature range of interest are mainly based on the X-ray analysis
of Flewitt \etal for the Nb-rich part \cite{Flewitt1972},
and on the experimental data (X-ray, resistivity, microscopical examination and dilatometry) 
of Van Effenterre for the Zr-rich part \cite{VanEffenterre1972}. 
However, the solubility of Zr in the $\beta$ phase resulting from this Calphad assessment \cite{Guillermet1991}
is lower than the experimental one reported by Van Effenterre \cite{VanEffenterre1972}. A revised Calphad assessment recently performed by Lafaye \cite{Lafaye2017} lead to the same solubility limits as Guillermet in both $\beta$-Nb and $\alpha$-Zr phases.

In the past decades, much progress has been made in the \abinit theory of phase stability \cite{DeFontaine1994,Ducastelle1991}. 
Knowing only atomic numbers, electronic structure calculations  are commonly employed 
to predict the structural and thermodynamic properties of a wide range of materials.
The purpose of this work is to use such an \abinit approach to complement experimental observations 
in Zr-Nb alloys, and to improve our understanding of the phase diagram of the binary system.
First-principles calculations relying on the density functional theory
are employed to investigate the thermodynamic properties of the Zr-Nb system.
First, the alloy energetics are evaluated at 0\,K and the results used 
in a simple thermodynamic model, where only configurational entropy is considered,
to determine the corresponding miscibility gap.
Then, as electronic excitations and lattice vibrations have been shown to influence the relative stability 
of the different phases in several systems \cite{Ozolins2001,Asta2001,Clouet2002,Burton2006},
we calculate these contributions to the alloy free energies for both $\alpha$ and $\beta$ solid solutions. 
Finally, the solubility limits predicted from the thermodynamic model accounting for all the free energy contributions
are compared to experimental data.

\section{Details of \abinit calculations \label{sec:method}}

\subsection{DFT parameters}
\label{subsec:general}

The electronic-structure calculations presented hereafter are based on the density functional theory (DFT) in the generalized gradient approximation with the exchange-correlation functional proposed by Perdew \etal \cite{Perdew1996}.
The core electrons are replaced with ultrasoft pseudo-potentials of Vanderbilt type, 
with 4s and 4p semi-core electrons included in the valence state both for Nb and Zr. 
The \pwscf code of the Quantum Espresso package~\cite{Giannozzi2009} is employed, using a plane-wave basis set with an energy cutoff of 680\,eV to describe the valence electrons. Such a high value is required to obtain well-converged phonon dispersion spectra (\cf \ref{sec:annex2}). 
A Monkhorst-Pack $\vec{k}$-points grid is used for integration in the reciprocal space,
with a $24\times24\times24$ and a $24\times24\times16$ mesh in bcc Nb and hcp Zr conventional cells, respectively, and an equivalent density of $\vec{k}$-points for the supercells.
The electronic density of states is broadened with the Methfessel-Paxton function with a smearing of 0.2\,eV. 

Our calculations are done at constant volume, thus relaxing only the atomic positions. 
Bcc and hcp supercell dimensions are fixed to the equilibrium lattice parameters of Nb and Zr.
Based upon test calculations performed with different $\vec{k}$ point meshes, plane wave cutoffs and electronic smearing, 
solution energies are estimated to be converged to within 2\,meV/atom, lattice parameters to within a fraction of a percent and electronic entropies to within $0.05\ k$/atom.

Equilibrium bulk properties of $\beta$-Nb and $\alpha$-Zr have been computed and compared to available experimental data to validate our \abinit approach. The obtained values, listed in Table \ref{tab:method}, are close to experimental data. 
The vacancy formation energies also agree with experiments \cite{Maier1979,Hood1984,Hood1986}, 
and with previous \abinit calculations \cite{Xin2009,Varvenne2014}.

\subsection{Solution energy}
\label{subsec:sol_energy}

Solution energies are obtained using single impurity calculations as well as special quasi-random structures (SQS). In single impurity calculations, we simply have one substitutional Zr solute atom in a $\beta$-Nb supercell and one Nb solute atom in an $\alpha$-Zr supercell. The SQS are used to create supercells which are good representative configurations of a random solid solution for a given concentration~\cite{Zunger1990}. They are generated using the code \mcsqs from the \atat package~\cite{vandeWalle2013}.

\subsection{Vibrational excitations}
\label{subsec:vib}

Harmonic vibrational spectra are computed using the density-functional perturbation theory (DFPT) \cite{deGironcoli1995} implemented in the Quantum Espresso package \cite{Giannozzi2009}. 
The same plane wave energy cutoff and $\vec{k}$-point sampling are used as for total energy calculations
(\S\ref{subsec:general}).
Test calculations performed on impurity supercells for modes near $\vec{q}=\Gamma$ wave vectors 
show that these parameters ensure convergence of phonon frequencies to within 0.05\,THz.
Dynamical matrices are calculated on a regular grid of $\vec{q}$ wave vectors.
$12\times12\times12$ and $6\times6\times5$ meshes are used
for the bcc Nb and hcp Zr elementary cells,
and equivalent grids with the same density of $\vec{q}$-points for larger supercells.
A denser grid has to be used for calculations in the $\beta$ Nb-rich phase 
because of the non zero values of the force constants up to the $13^{\rm th}$ nearest neighbors
in bcc Nb \cite{Nakagawa1963,Powell1968} (see \ref{sec:annex2}).
The dynamical matrices obtained by DFPT on a regular grid 
are then Fourier transformed to real space to yield interatomic force constants.
These force constants are subsequently used to interpolate phonon dispersions to arbitrary wave vectors $\vec{q}$. 
Finally, vibrational free energies are calculated with Eq.~\eqref{eq:Fvib} using a regular grid of $90\times90\times90$ $\vec{q}$-points for the bcc Nb-rich phase
and $90\times90\times60$ for the hcp Zr-rich phase.

\begin{table}[tb]
	\caption{Bulk properties of bcc Nb and hcp Zr compared to experimental data at 4.2\,K:
	lattice parameter $a$ (\AA), 
	relaxed elastic moduli $C_{ij}$ (GPa),
	bulk modulus $B$ (GPa),
	and vacancy formation energy $E^{\rm f}_{\rm V}$ (eV).}
	\label{tab:method}
	\subfloat[bcc Nb]{
		\begin{tabular}{lrr}
			\hline
					&\pwscf	&Exp.\footnotemark[1]\\
			\hline
			$a$		& 3.309 & 3.303  \\
			$C_{11}$	& 242.0 & 252.7  \\
			$C_{12}$	& 134.6 & 133.2  \\
			$C_{44}$	& 16.0	&  31.0  \\
			$B$		& 170.4 & 173.0  \\
			$E^{\rm f}_{\rm V}$	&  2.7	& $2.6\pm0.3$ \\%quenching, PAS, Doppler broadening
			\hline
		\end{tabular}
	}
	\hfill
	\subfloat[hcp Zr]{
		\begin{tabular}{lrr}
			\hline
					      & \pwscf &Exp.\footnotemark[1]\\
			\hline
			$a$                   &      3.240 &      3.232   \\
			$c/a$                 &      1.600 &      1.603   \\
			$C_{11}$              & 148.8      & 155.4        \\
			$C_{12}$              &   58.3       &   67.2       \\
			$C_{13}$              &   65.2       &   64.6       \\
			$C_{33}$              & 163.0       & 172.5       \\
			$C_{44}$              &   25.6       &   36.3       \\
			$C_{66}$              &   45.3       &   44.1       \\
			$B$                        &   93.1       &   97.3       \\
			$E^{\rm f}_{\rm V}$                 &     2.0        &  $>$~1.5        \\
			\hline
		\end{tabular}
	}\\
	{\footnotesize $^1$ Elastic moduli from Refs. \cite{Carroll1965} (Nb) and \cite{Fisher1964} (Zr),
	lattice parameters from \cite{Roberge1974} (Nb) and \cite{Goldak1966} (Zr),
	and vacancy formation energies from \cite{Maier1979} (Nb) and \cite{Hood1984,Hood1986} (Zr).}
\end{table}

\section{Solution and interaction energy \label{sec:energies}}

We are interested in the thermodynamic properties of both bcc Nb-rich and hcp Zr-rich solid solutions.
To define the free energies of these two phases, we first calculate the corresponding solution energies.
We then calculate the interaction energy between solute atoms 
to check that both solid solutions can be considered as simple unmixing alloys.

\subsection{Solution energy~\label{subsec:solubility_energy}}

The solution energy $\Omega^{\Phi}_{\rmix}$ of a substitutional B atom in a matrix of A atoms
with the structure $\Phi=\{\beta, \alpha\}$ 
is obtained by considering the excess energy per atom of a supercell containing $N$ sites 
with A and B atoms
\begin{align*}
\Delta E[ \rA_{N-M} \rB_{M}] = & \left\{ \; E[ \rA_{N-M} \rB_{M}] - (N-M)  \, E_{\rm{ref}}[ \rA ]   \right. \\ 
                                    & \left.   - \, M \, E_{\rm{ref}}[\rB] \;  \right\} / \, N ,
\end{align*}
where $E\left[ \rA_{N-M} \rB_{M} \right]$ is the energy of the supercell with $(N-M)$ A atoms and $M$ B atoms,
$E_{\rm{ref}}\left[ \rA  \right]=E[A_{N}]/N$ the reference energy of the matrix calculated with the same supercell,
and $E_{\rm{ref}}\left[ \rB  \right]$ the energy per atom of B in its equilibrium reference state,
i.e. bcc for $\beta$-Nb and hcp for $\alpha$-Zr.
This excess energy is just the solution energy weighted by the solute concentration $x_{\rB} = M/N$
\begin{equation}
\Delta E[ \rA_{N-M} \rB_{M}] = x_{\rB} \; \Omega^{\Phi}_{\rmix}.
\label{eq:solution_energy}
\end{equation}

As mentioned in~\S\ref{subsec:sol_energy}, the supercells used for the different concentrations are built using two methods: single impurity calculations and SQS. The SQS allow us to check the influence of the solute local environment on the solution energy. Regardless of the method, the total number $N$ of atoms in each supercell goes from 8 to 250. 

\begin{figure}[bt]\centering
(a)\includegraphics[width=7.2cm]{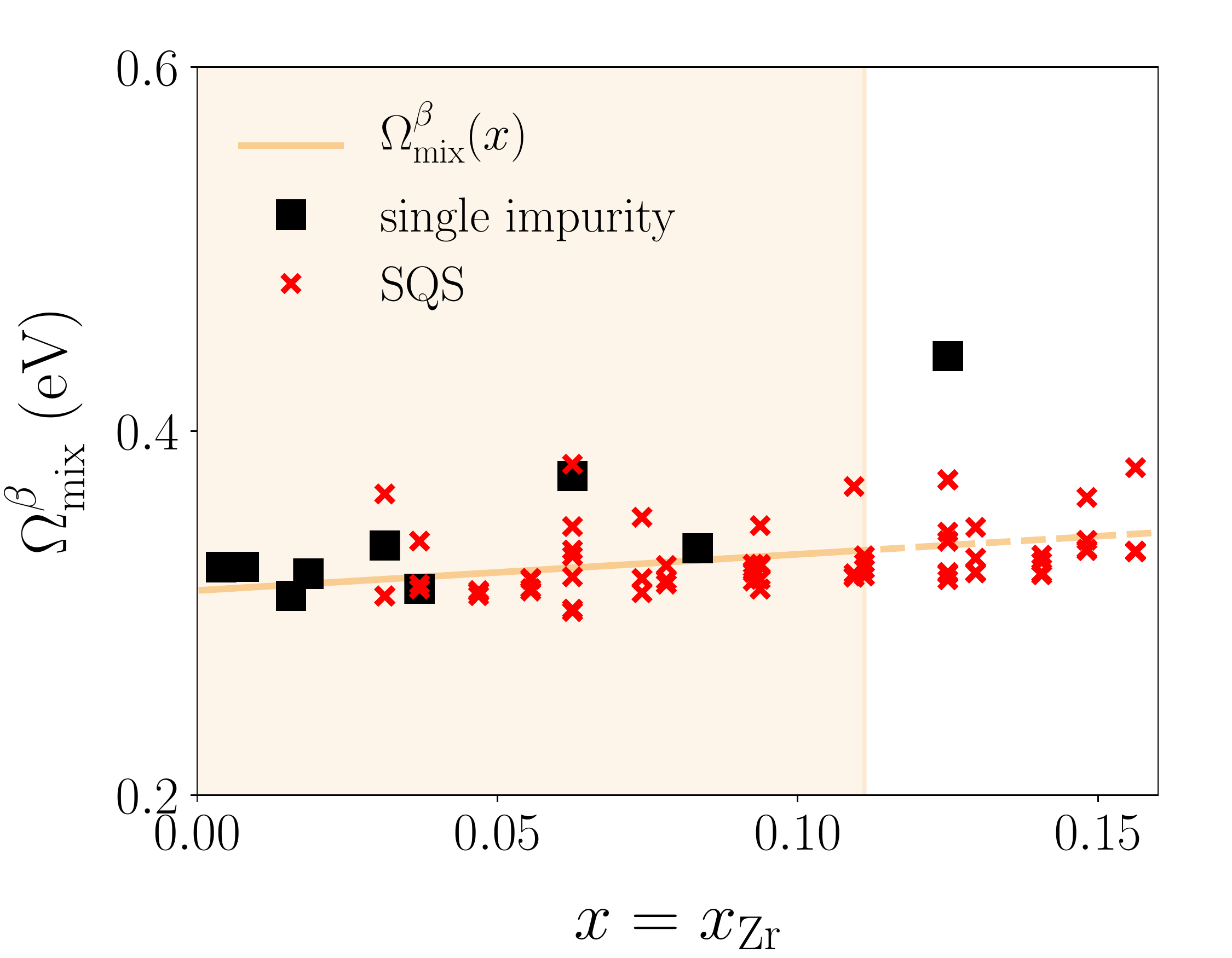}
(b)\includegraphics[width=7.2cm]{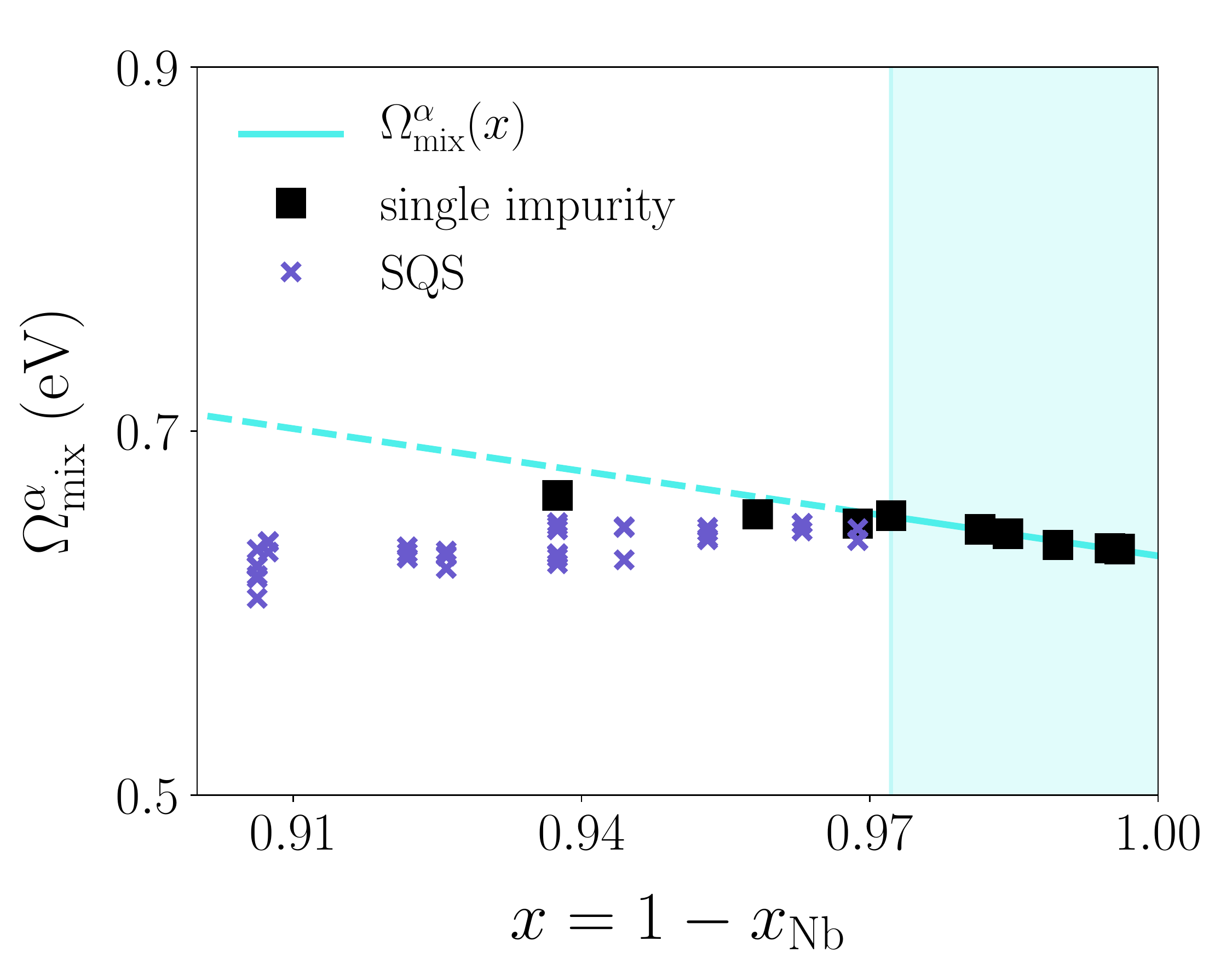}
	\caption{Solution energies $\Omega^{\Phi}_{\rmix}$ as a function of the atomic fraction $x=x_{\rZr}=1-x_{\rNb}$ for (a) bcc Nb-rich and (b) hcp Zr-rich solid solutions.
	The shaded regions correspond to the concentration range used to fit $\Omega^{\Phi}_{\rmix}(x)$ on the \abinit data (continuous line).}
	\label{fig_1} 
\end{figure}

The results are represented in Figure \ref{fig_1}.
The positive values agree with the strong unmixing tendency of the Zr-Nb alloy.
Regarding the Nb-rich part (Fig. \ref{fig_1}a), single impurity calculations lead to oscillations of the solution energy. 
The supercells are probably too small to avoid any short-range order effects. 
The SQS calculations, on the other hand, show a smoother variation with the solute concentration.
For the Zr-rich part, single impurity calculations are sufficient to obtain solution energies within the concentration range explored by experimental data. 

Defining $x=x_{\rZr}=1-x_{\rNb}$, the solution energy $\Omega^{\Phi}_{\rmix}(x)$ has been interpolated linearly 
as a function of concentration (Fig.~\ref{fig_1}), 
restricting the fit to the lower concentrations, 
$x_{\rZr}\leq0.11$ for the $\beta$ phase and $x_{\rNb}\leq0.03$ for the $\alpha$ phase.
Such a linear variation describes reasonably well the solution energy 
in the concentration range where the solid solutions are stable. 
Moreover, we will show hereafter that the solution energy variations with solute concentration 
are actually small enough to have a negligible impact on the solubility limits. Therefore, we choose to neglect higher order terms in the concentration dependence of the solution energy.
The obtained parameters are given in Table~\ref{tab:Esol}. 

The solution energy for infinite dilution is  
$\Omega^{\beta}_{\rmix}(x\rightarrow0) = 0.312$\,eV for Zr impurity in $\beta$-Nb,
and $\Omega^{\alpha}_{\rmix}(x\rightarrow1) = 0.632$\,eV for Nb impurity in $\alpha$-Zr,
in agreement with previous \abinit calculations ($0.61$\,eV for Domain \cite{Domain2006}
and $0.68$\,eV for Xin \etal \cite{Xin2009}).
Our results are also consistent with the asymmetry observed in the phase diagram,
as the high value of $\Omega^{\alpha}_{\rmix}$ explains the very low solubility of niobium in hcp zirconium,
whereas the lower value of $\Omega^{\beta}_{\rmix}$ corresponds to a higher solubility of zirconium in bcc niobium. 
Moreover, Guillermet \cite{Guillermet1991} and Lafaye \cite{Lafaye2017} also assessed that the solution energy of the hcp phase of Zr-Nb alloys 
is larger than the one of the bcc phase by a factor of $2$.

\begin{table}
	\centering
	\caption{Contributions to the solution free energies:
		solution energy $\Omega^{\Phi}_{\rmix}$ at 0\,K,
		as well as vibrational $\Omega^{\Phi}_{\rvib}$ 
		and electronic $\Omega^{\Phi}_{\rel}$ contributions.}
	\label{tab:Esol}
	\subfloat[$\beta$ Nb-rich solid solution]{
	\begin{tabular}{ccrcr}
		\hline
		$\Omega^{\beta}_{\rmix}$ 		& (eV)		& $ 0.312$ & $ +$ & $ 0.199 \, x_{\rZr} $ \\
		$\Omega^{\beta}_{\rvib}/kT$		&		& $-0.41$  &      &                       \\
		$\Omega^{\beta}_{\rel}/(\pi k T)^2$ 	& (eV$^{-1})$	& $-0.13$  & $ -$ & $ 0.86  \, x_{\rZr} $ \\
		\hline
	\end{tabular}}

	\subfloat[$\alpha$ Zr-rich solid solution]{
	\begin{tabular}{ccrcr}
		\hline
		$\Omega^{\alpha}_{\rmix}$ 		& (eV)		& $ 0.632$ & $ +$ & $ 0.776 \, x_{\rNb} $ \\
		$\Omega^{\alpha}_{\rvib}/kT$		&		& $-4.59$  & $ -$ & $10.4   \, x_{\rNb} $ \\
		$\Omega^{\alpha}_{\rel}/(\pi k T)^2$ 	& (eV$^{-1})$	& $-0.71$  & $ -$ & $ 0.44  \, x_{\rNb} $ \\
		\hline
	\end{tabular}}
\end{table}

\subsection{Binding energies \label{subsec:binding_energies}}

The binding energies between two Zr (Nb) solute atoms in a bcc Nb (hcp Zr) matrix have been determined 
in order to check the type and range of interactions in the system.
The binding energy between two B impurities lying on $n^{\textrm{th}}$ nearest-neighbor sites in an A matrix is given by
\begin{equation}
E^{b}_{\rB-\rB} =  2  E\left[\rA_{N-1} \rB \right] 
			- E\left[ \rA_N \right]
			- E\left[ \rA_{N-2} \rB_{2} \right],
\label{eq:eq_3}
\end{equation}
where $E\left[ \rA_{N-2} \rB_{2} \right]$ is the energy of the supercell containing $N-2$ A atoms and one B$-$B pair of solute atoms at the nearest-neighbor distance $d_n$, 
$E\left[ \rA_{N-1} \rB \right]$ the energy of the same supercell with a single impurity atom,
and $E\left[ \rA_N  \right] $ the energy of the supercell with only $A$ solvent atoms. 
Positive values correspond to attractive interactions.
The calculations are made for $128$-atom and $96$-atom supercells for the bcc Nb and hcp Zr matrix, respectively.

\begin{figure}[h!]\centering
	(a)\includegraphics[width=7.5cm]{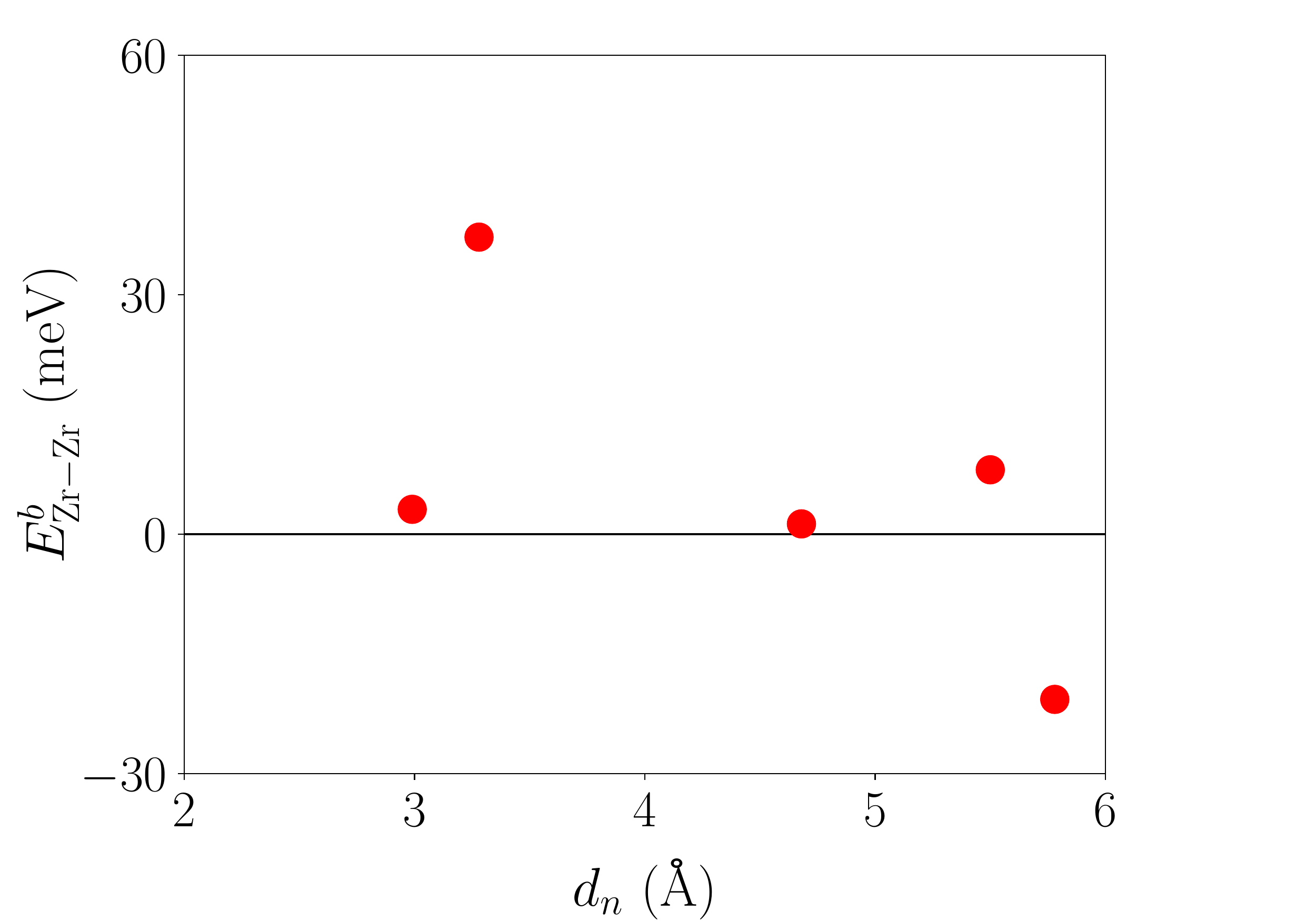}
	(b)\includegraphics[width=7.5cm]{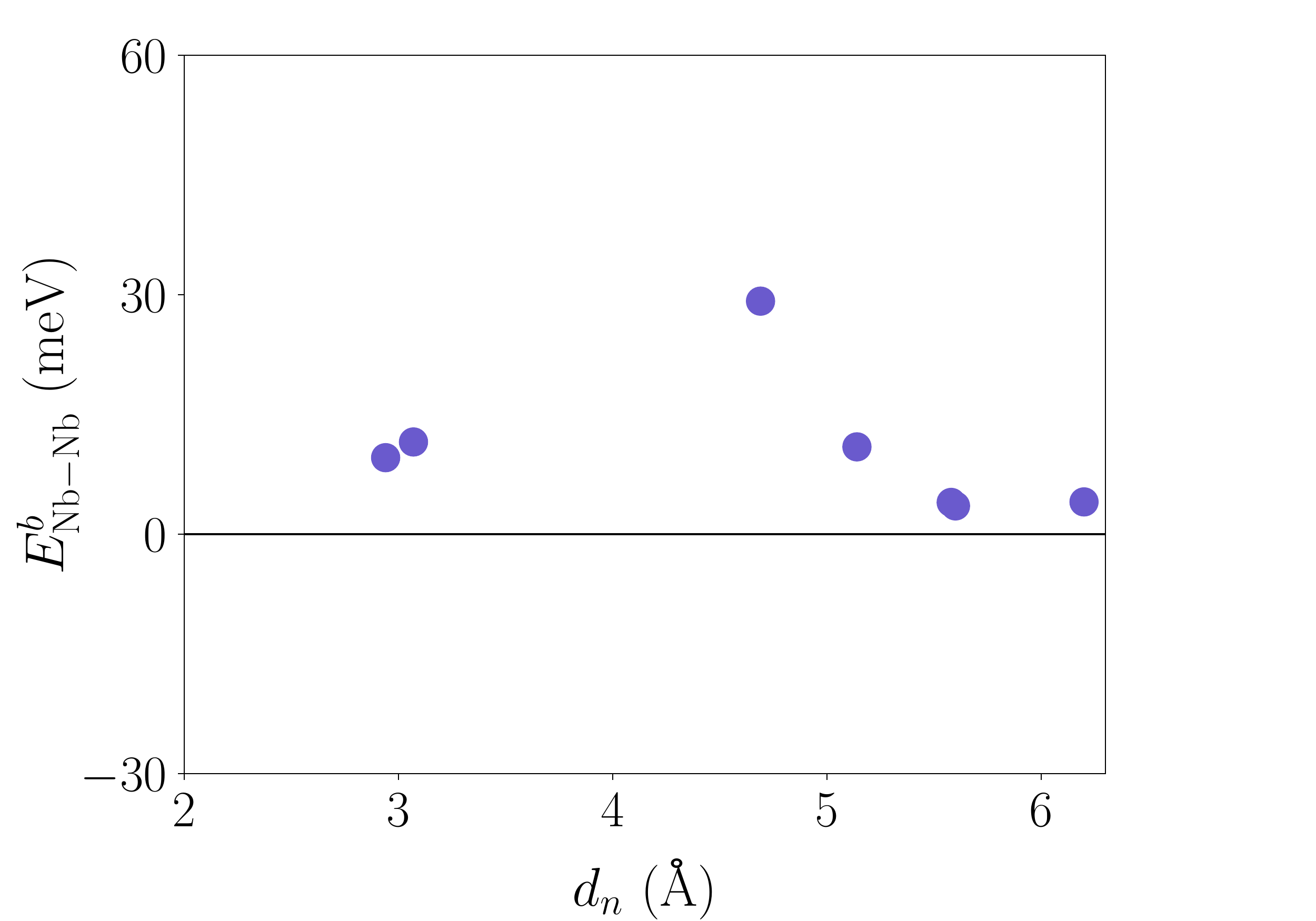}
	\caption{Binding energies between (a) Zr solute atoms in a bcc Nb matrix
	and (b) Nb solute atoms in an hcp Zr matrix. 
	The binding energies are shown as a function of the separation distance $d$ 
	after relaxation.}
	\label{fig_2} 
\end{figure}

In both cases, the unmixing tendency results in slightly positive binding energies between solute atoms (points in Fig.~\ref{fig_2}).
Overall the interactions are weak with values below 40\,meV. 
In the bcc Nb matrix, the only relevant interaction corresponds to the second nearest neighbor position, 
as the two Zr solute atoms almost do not interact when they are first nearest neighbors or for larger separation distances. 
A slightly negative binding energy is obtained for the largest separation distance, which probably arises from the supercell being too small to keep on neglecting interactions with the periodic images of the pair of solute atoms.
A similar, but less extreme, behavior is observed for the pair of Nb solute atoms in the Zr hcp matrix, 
as the binding energy is lower between solute atoms situated in position of first- than in position of second-nearest neighbors. Beyond second nearest neighbor,
the binding energy smoothly decreases with an increasing separation distance.
Both solid solutions appear as simple unmixing systems, with no ordering tendency. 
Therefore, only these two phases need to be considered when calculating the phase diagram \cite{Ducastelle1991},
with simple thermodynamic approximations used for their free energy functionals (\S\ref{sec:thermo_model}). 

\section{Thermodynamic model \label{sec:thermo_model}}

With the energetics properties assessed from DFT calculations, we can build a thermodynamic model for the Zr-Nb system and get a better understanding of the Zr-Nb solid state phase diagram for temperatures below $890$\,K. 
As both the bcc and hcp solid solutions are simple unmixing systems without any pronounced ordering tendency,
we use the Bragg-Williams approximation which neglects all correlations between different lattice sites.
The excess free energies of the two phases are then simply given by
\begin{align}
	\Delta F^{\beta}_{\rmix}(x,T)  &= x \, \Omega^{\beta}_{\rmix}(x) \nonumber \\
		&+ k T \left[ x \ln{(x)} + (1-x) \ln{(1-x)} \right]  \nonumber, \\
	\Delta F^{\alpha}_{\rmix}(x,T) &= (1-x) \, \Omega^{\alpha}_{\rmix}(x) \nonumber    \\ 
		&+ k T \left[ x \ln{(x)} + (1-x) \ln{(1-x)} \right]  ,
\label{eq:Fmix}
\end{align}
where $k$ is the Boltzmann constant.
The second term on the right hand side corresponds to the configurational entropy of the binary solid solutions.
All quantities are defined with respect to Zr atomic fraction $x=x_{\rZr}=1-x_{\rNb}$. 

Equilibrium is given by numerical minimization of $F^{\beta}(x,T)$ and $F^{\alpha}(x,T)$ using the common tangent construction. 
If one neglects the concentration dependence of the solution energy, 
considering their values for infinite dilution $\Omega^{\beta}_{\rmix}(x\to0)$ and $\Omega^{\alpha}_{\rmix}(x\to1)$,
analytical expressions of the solubility limits can also be obtained,
\begin{align}
	x^{\beta} &=  \frac{ 1 - \exp{\left( - \frac{ \Omega^{\alpha}_{\rmix}}{kT} \right)} }
	{ 1 -\exp{\left( - \frac{ \Omega^{\beta}_{\rmix}+\Omega^{\alpha}_{\rmix} }{kT} \right)} }  
	\exp{\left( - \frac{ \Omega^{\beta}_{\rmix}}{kT} \right)},                      \nonumber \\
	x^{\alpha}&= 1 - \frac{1 - \exp{\left( - \frac{\Omega^{\beta}_{\rmix}}{kT} \right)} }{ 1 -\exp{\left( - \frac{ \Omega^{\beta}_{\rmix}+\Omega^{\alpha}_{\rmix} }{kT} \right)} }          
	\exp{\left( - \frac{ \Omega^{\alpha}_{\rmix}}{kT} \right)}.
\label{eq:eq_7}
\end{align}
One can further approximate these expressions, by keeping only the leading terms,
\begin{align}
	x^{\beta}  &=     \exp{\left( - \frac{ \Omega^{\beta}_{\rmix}} {k\,T} \right)},                   \nonumber \\
	x^{\alpha} &= 1 - \exp{\left( - \frac{ \Omega^{\alpha}_{\rmix}}{k\,T} \right)}.
	\label{eq:solubility_mix_dilute}
\end{align}
This approximation corresponds to the equilibrium between the $\beta$ solid solution and the $\alpha$ pure phase, 
and vice versa.

\begin{figure}[!bt]\centering
	(a)\includegraphics[width=1.\linewidth]{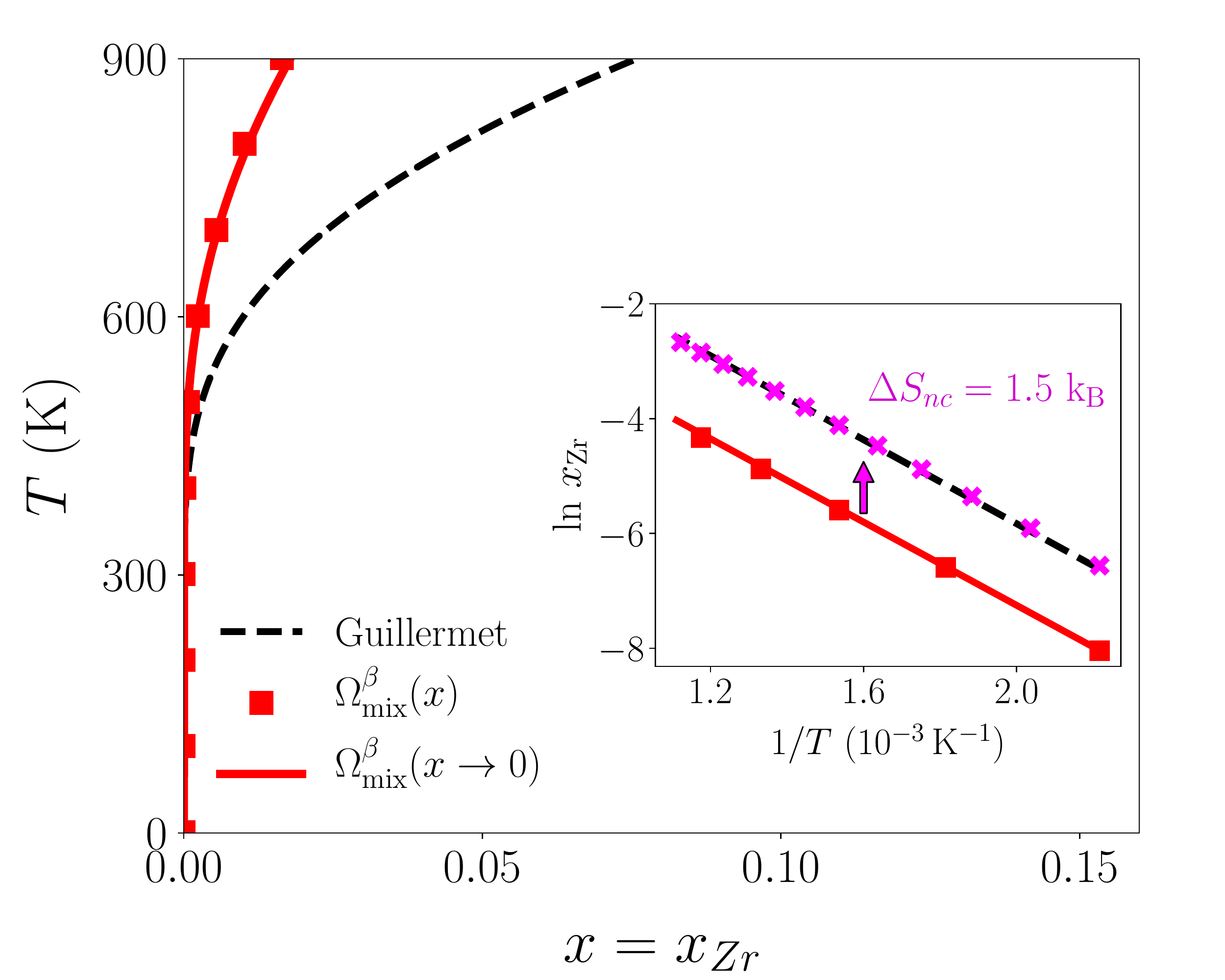}
	(b)\includegraphics[width=1.\linewidth]{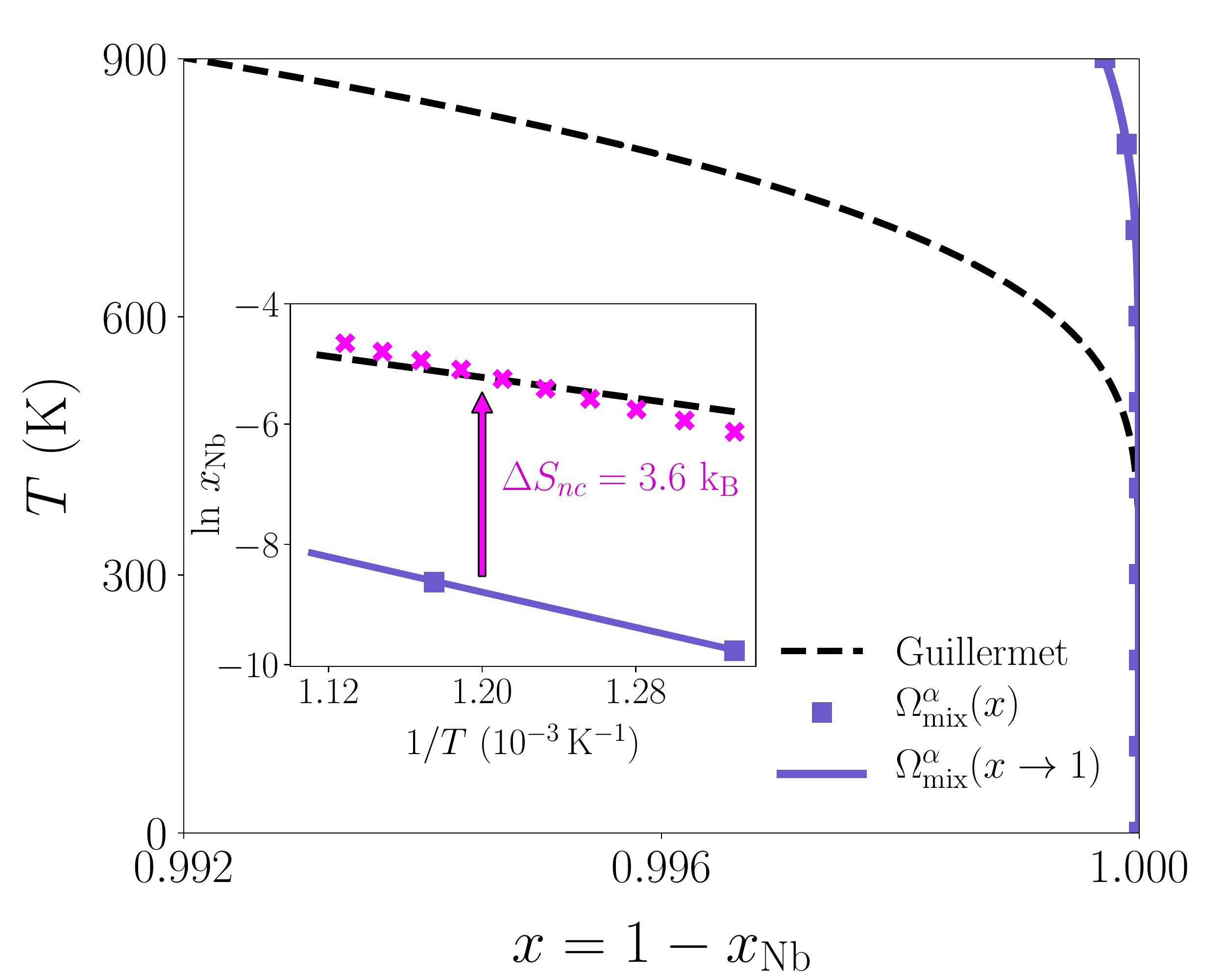}
	\caption{Solubility limits (a) $x^{\beta}$ in the bcc Nb-rich phase
	and (b) $x^{\alpha}$ in the hcp Zr-rich phase
	obtained with temperature independent solution energies. 
	The solubility limits given by numerical minimization, 
	taking full account of the concentration dependence of the solution energy,
	are shown by colored squares 
	and their dilute limit approximations (Eq. \ref{eq:solubility_mix_dilute}) 
	by continuous lines. 
	They are compared to the Calphad assessment of Guillermet \cite{Guillermet1991}.
	Arrhenius plots are shown in insets, with the solubility limits obtained 
	when a non-configurational entropy $\Delta S_{\rm nc}$ is fitted (purple crosses).
	}
	\label{fig:solubility_mix} 
\end{figure}

These analytical expressions, which neglect the concentration dependence of the solution energy,
give solubility limits which fully agree with the ones obtained by numerical minimization 
of the free energies (Eq. \ref{eq:Fmix}) accounting for this dependence (Fig. \ref{fig:solubility_mix}). 
This is true even for the expressions given by Eq. \ref{eq:solubility_mix_dilute} which correspond to the dilute limits. 
It justifies a posteriori our choice of a simple energy model to describe the thermodynamics
of the solid solutions.

The previous expressions are obtained when accounting for solution energy and configurational entropy only, 
neglecting non-configurational contributions like electronic and vibrational free energies. 
Under this approximation, the predicted solubility limits do not exhibit a good agreement with experimental data \cite{Guillermet1991}, even though the asymmetry of the miscibility gap tends to be correct (Fig. \ref{fig:solubility_mix}).
However, the Arrhenius plots $\ln{(x)} = f(1/T)$ show that the predicted slope is correct 
(insets in Fig.~\ref{fig:solubility_mix}). 
As shown by Eq. \ref{eq:solubility_mix_dilute}, this slope corresponds to the solution energy. 
Therefore, the solution energies deduced from \abinit calculations are in the expected range, 
but a non-configurational entropy $\Delta S_{\rm nc}$ needs to be accounted for to reproduce the experimental solubility limits. 
Assuming a constant entropy contribution to the solution energy, 
the straight line in these Arrhenius plots will now be given by
\begin{equation*}
	\ln{\left( x^{\beta} \right)} = - \frac{ \Omega^{\beta}_{\rmix}} {k\,T} + \frac{ \Delta S^{\beta}_{\rm nc}}{k}.
\end{equation*}
Estimating the values of $\Delta S_{\rm nc}$ on the experimental data leads to
$1.5\,k$ for the solution free energy of Zr in the $\beta$ phase
and a much higher one for Nb in the $\alpha$ phase with $3.6\,k$ (insets in Fig.~\ref{fig:solubility_mix}).
Next, these contributions are assessed using \abinit calculations.

\section{Finite temperature excitations \label{sec:temperature_prop}}

We now incorporate the contributions of electronic excitations and lattice vibrations 
in the free energies of the solid solutions. 
We use an adiabatic approximation \cite{DeFontaine1994}, 
assuming a separation of time scales, with a much faster one for electronic excitations and lattice vibrations than for configuration changes.
The electronic and vibrational free energies can then be defined for each alloy configuration. 
Under Bragg-Williams approximation, thus neglecting any short range order, 
the configurations of the solid solution are simply identified by their concentration, 
leading to the free energy
\begin{equation}
F^{\Phi}(x,T) = F^{\Phi}_{\rmix}(x,T) + F^{\Phi}_{\rel}(x,T)+ F^{\Phi}_{\rvib}(x,T),
\label{eq:eq_4}
\end{equation}
where $F^{\Phi}_{\rmix}$, $F^{\Phi}_{\rel}$ and  $F^{\Phi}_{\rvib}$ are the mixing, electronic and vibrational free energies
of the phases $\Phi=\alpha$ or $\beta$.
We define a contribution of electronic excitations and lattice vibrations 
to the solution free energy, respectively $\Omega^{\Phi}_{\rel}$ and $\Omega^{\Phi}_{\rvib}$, such as for the solution energy (Eq. \ref{eq:solution_energy}).
The corresponding excess free energies are then given by this contribution weighted by the solute concentration.

\subsection{Electronic free energy \label{subsec:elec_free_energy}}

At finite temperature, electrons close to the Fermi level can reach electronic states of higher energy 
following Fermi-Dirac distribution. 
We consider these electronic excitations in a static non vibrating lattice,
assuming a temperature independent electronic density of states (DOS), 
which is usually an accurate approximation \cite{Wolverton1995,Zhang2017}.
The energy and entropy entering the electronic free energy are then given by 
\begin{subequations}
	\label{eq:Fel}
	\begin{align}
		E_{\rm el}(T) = & \int_{-\infty}^{+\infty} E \; n(E) \left\{ f(E,T) - f(E,0) \right\} \md E , 
		\label{eq:Eel} 
		\\
		S_{\rm el}(T) = &-k  \int_{-\infty}^{+\infty} n(E) \left\{ f(E,T) \ln{\left[ f(E,T) \right]}  \right. \nonumber \\ 
		& \left. + \left[1- f(E,T)\right] \ln{\left[1- f(E,T) \right]}  \right\} \md E ,
		\label{eq:Sel}
	\end{align}
\end{subequations}
where $f(E,T) = 1/\left\{ \exp{\left[ (E-\mu(T))/kT \right]} + 1 \right\}$ is the Fermi-Dirac distribution. 
The chemical potential $\mu(T)$ is  defined through the conservation of the total number of conduction electrons 
$N_e = \int n(E) \, f(E,T) \, \md E$.
This chemical potential is equal to the Fermi energy $E_{\rm F}$ at 0\,K,
and needs to be reevaluated at finite temperature because of the variations of the electronic DOS around the Fermi level \cite{Ashcroft1976}.
The second term on the right-hand side of equation \eqref{eq:Eel} removes the electronic energy contribution at 0\,K already accounted for in $\Omega^{\Phi}_{\rmix}$~(\S\ref{subsec:solubility_energy}). 
In the present work, instead of using Eq. \eqref{eq:Eel} to calculate the energy term, 
we determine it from $E_{\rel}(T_0)=\int_0^{T_0} T\, \left( \frac{\partial S_{\rel}}{\partial T}\right) \md T$,
which gives a better precision \cite{Satta1998}. 

\begin{figure}[!bth]
	\centering
	(a)\includegraphics[width=7.6cm]{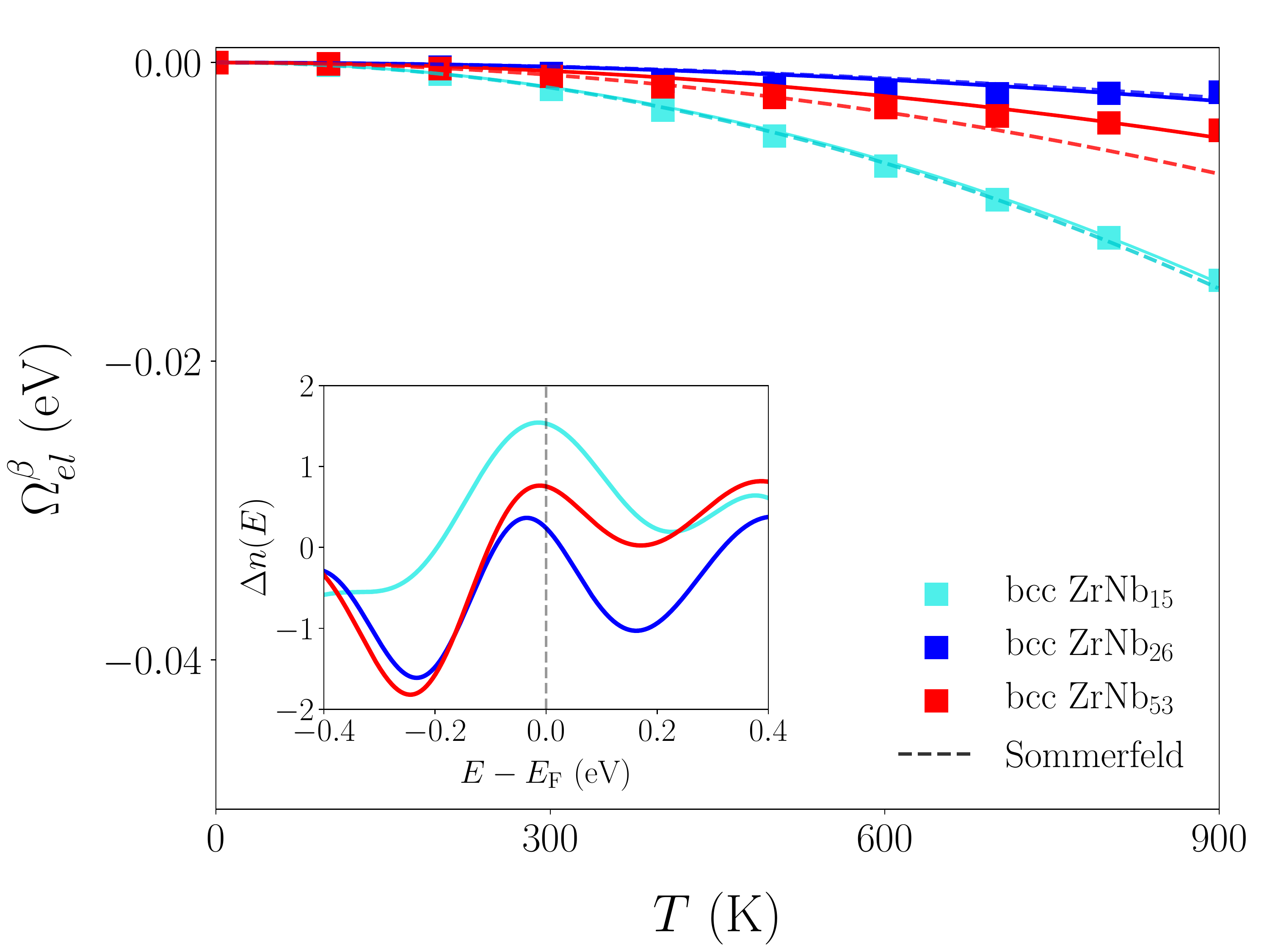}
	(b)\includegraphics[width=7.6cm]{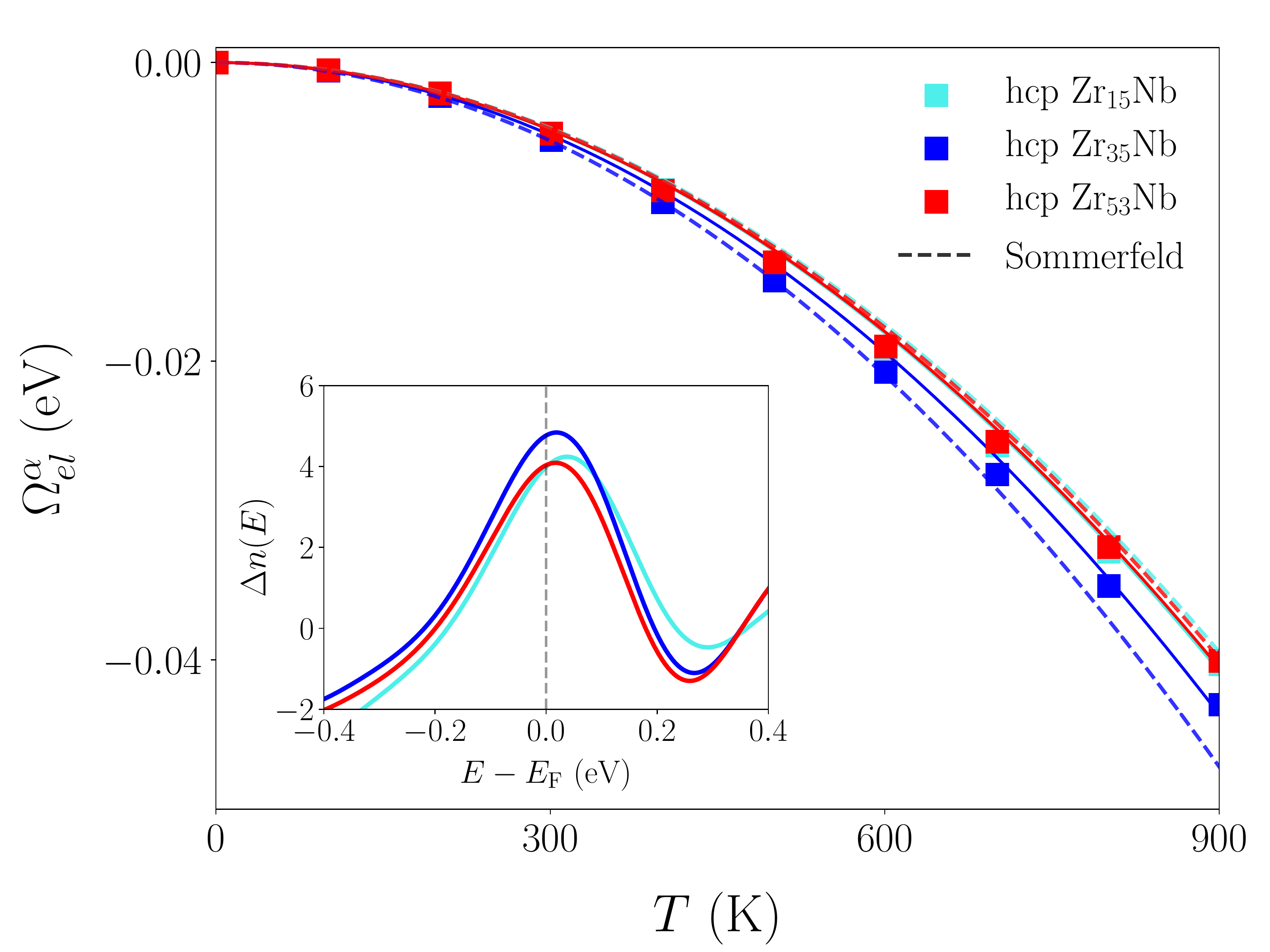}
	\caption{Contribution of electronic excitations to the solution free energy 
	for (a) bcc ZrNb$_{N-1}$ and (b) hcp Zr$_{N-1}$Nb compounds.
	The symbols are the results of the numerical integration (Eq. \eqref{eq:Fel})
	and the lines correspond to Sommerfeld model with an electronic DOS $\Delta n(E_{\rm F})$
	given by \abinit calculations (dashed lines)
	or fitted to reproduce results of the numerical integration (solid lines).
	The corresponding differences of the electronic DOS $\Delta n(E)$ entering in the definition 
	of the excess free energy are represented in the insets.
	}
	\label{fig:Fel} 
\end{figure}

Excess electronic free energies are evaluated for all the single impurity calculations 
described in section \S\ref{subsec:solubility_energy}.
The results are shown in Fig. \ref{fig:Fel} for three such calculations
bcc ZrNb$_{N-1}$ with $N=16$, 27 and 54 in the $\beta$ phases
and hcp Zr$_{N-1}$Nb with $N=16$, 36 and 54 in the $\alpha$ phase.
The corresponding differences $\Delta n(E)$ between the electronic DOS of the compounds
and the composition weighted DOS of the pure phases are represented in the insets. 
A special $\vec{k}$-point mesh was used for the calculation bcc Nb$_{26}$Zr as described in \ref{sec:annex1}.
For both Zr and Nb impurities calculations, the electronic contribution to the solution free energy is small
with absolute values not exceeding 50\,meV at the highest temperature. 
This contribution is higher for Nb impurity than for Zr impurity 
because of larger differences of the electronic DOS $\Delta n(E_{\rm F})$ at the Fermi level.

\begin{figure*}[!bth]
	\centering
	(a)\includegraphics[width=7.5cm]{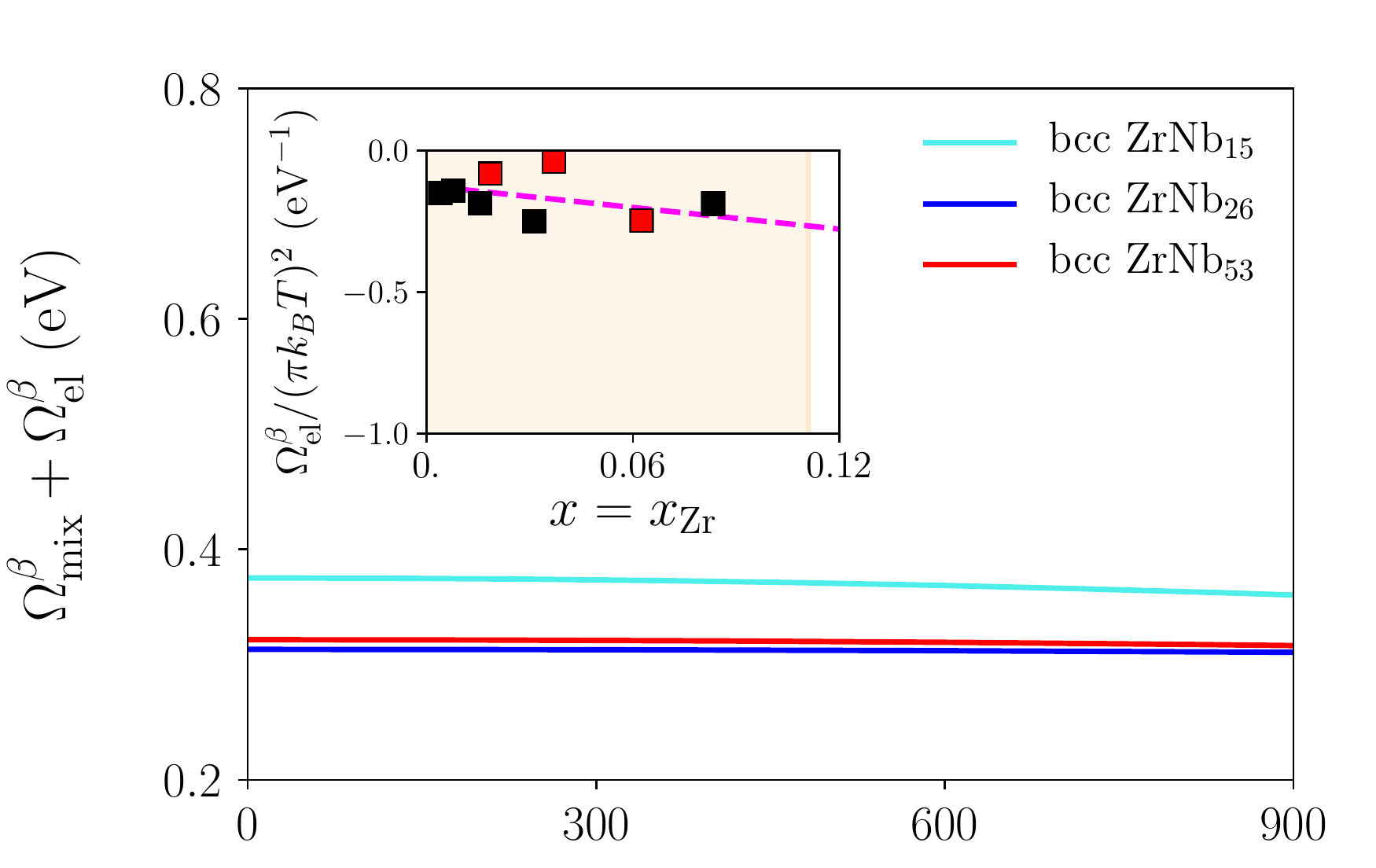}
	(b)\includegraphics[width=7.5cm]{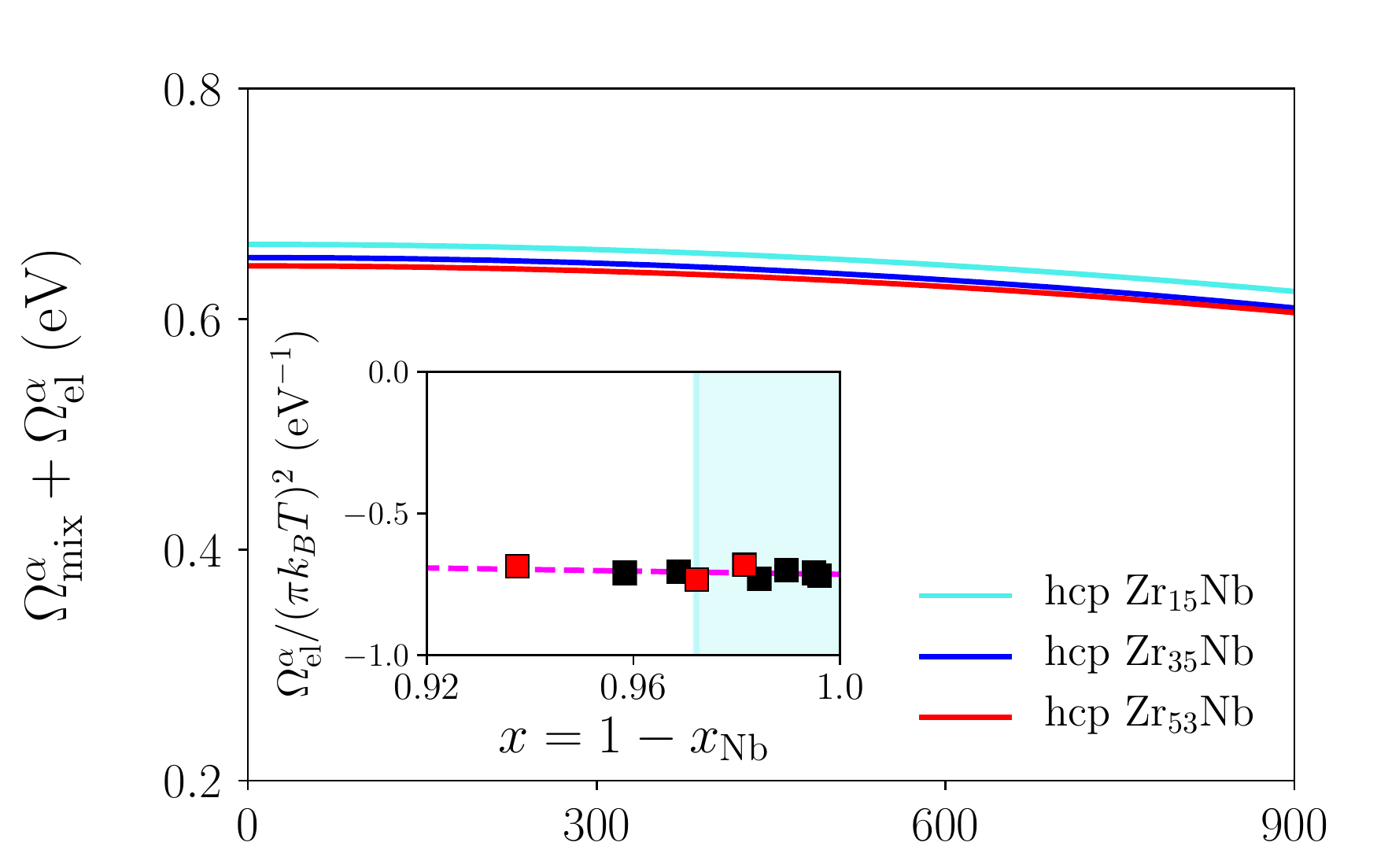}
	(c)\includegraphics[width=7.5cm]{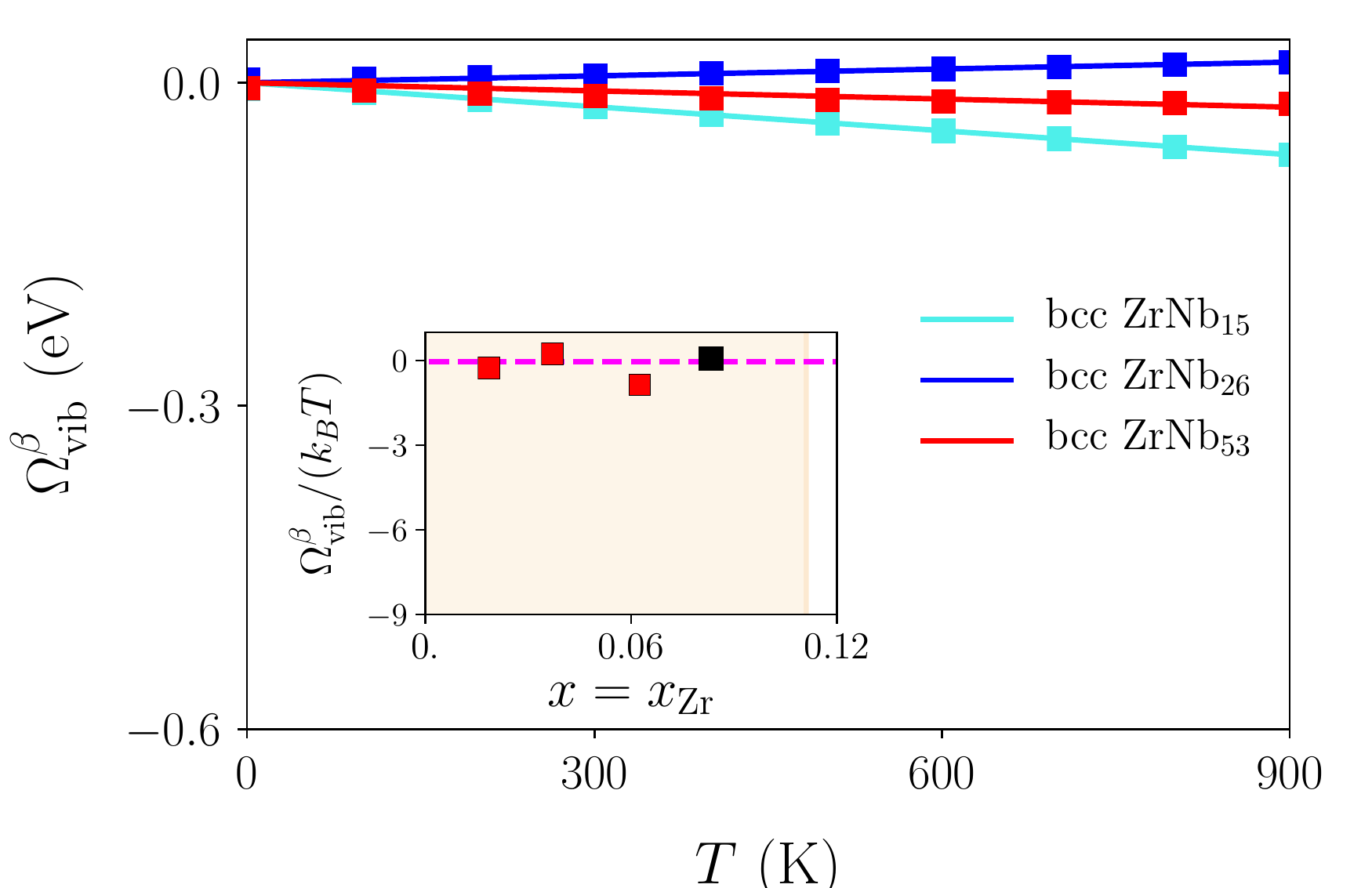}
	(d)\includegraphics[width=7.5cm]{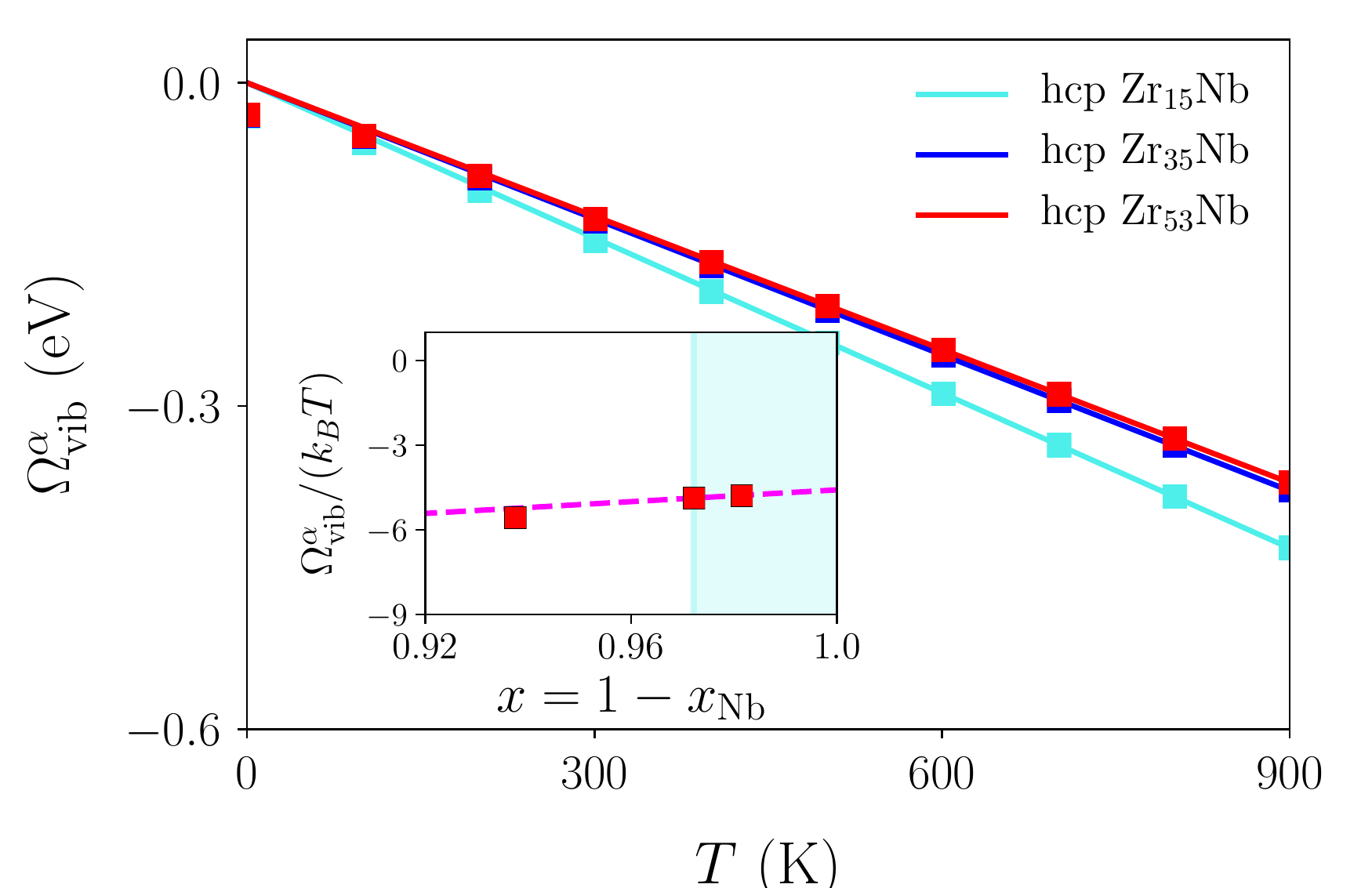}
	\caption{Contributions to the solution free energies for (a) and (c) bcc ZrNb$_{N-1}$,
	(b) and (d) hcp Zr$_{N-1}$Nb compounds.
	The contributions $\Omega^{\Phi}_{\rel}(T)$ of electronic excitations are added
	to the solution energy $\Omega^{\Phi}_{\rmix}(T)$ at 0\,K in (a) and (b),
	and the vibrational contribution $\Omega^{\Phi}_{\rvib}(T)$ is shown in (c) and (d).
	The insets show the corresponding contributions normalized by their temperature dependence 
	as a function of the atomic fraction. 
	Symbols are direct results from \abinit calculations and dashed lines correspond to the thermodynamic model
	(Table \ref{tab:Esol}).
	}
	\label{fig_5} 
\end{figure*}

Under the same assumption of a temperature independent electronic DOS,
the Sommerfeld model predicts that the electronic free energy is given by 
$F_{\rm el}(T) = -(\pi k T)^2 n(E_{\rm F}) / 6$ \cite{Ashcroft1976,Grimwall1986,Wolverton1995}. 
The corresponding excess free energy $\Delta F^{\Phi}_{\rm el}(T)$,
and thus the contribution $\Omega^{\Phi}_{\rm el}$ to the solution free energy,
show the same quadratic variation with temperature 
with a prefactor depending on $\Delta n(E_{\rm F})$. Sommerfeld model is in good agreement with the results obtained by numerical integration 
of Eq. \eqref{eq:Fel} (Fig.~\ref{fig:Fel}).
To improve the agreement and obtain a quantitative analytical expression of this electronic contribution,
we fit the prefactor corresponding to the electronic DOS.
The contributions $\Omega^{\Phi}_{\rm el}(T)$ of electronic excitations, normalized by $(\pi k T)^2$,
are shown for all single impurity calculations in the insets of Fig. \ref{fig_5}. 
This electronic contribution is then interpolated linearly as a function of concentration 
to obtain an analytical expression entering the definition \eqref{eq:eq_4} of the solid solutions excess free energies.
Parameters of this linear interpolation are given in table \ref{tab:Esol}.
By adding this electronic contribution $\Omega^{\Phi}_{\rm el}$ to the solution energy $\Omega^{\Phi}_{\rm mix}$,
one now obtains a solution free energy which depends on temperature (Fig. \ref{fig_5}). 
However, this temperature dependence remains small, 
thus showing that electronic excitations only marginally contribute to the excess free energies 
of the solid solutions.

~\\

\subsection{Vibrational free energy \label{subsec:vib_free_energy}}

Atomic vibrations are incorporated in the free energies within the harmonic approximation.
The vibrational free energy of a supercell with $N$ lattice sites is then given by
\begin{equation}
	F_{\rm vib}(T) =  \sum^{3N-3}_{\vec{q}, s} \left\{ \frac{\hbar \omega_{\vec{q}}^{s}}{2}  
	+  k T \ln{ \left[ 1 - \exp{\left( - \frac{ \hbar \omega_{\vec{q}}^{s} }{ k T } \right)} \right] } \right\},
	\label{eq:Fvib}
\end{equation}
where the sum runs on $3N-3$ modes out of the total number of vibration pulsations $\omega_{\vec{q}}^s$,
since three null eigenvalues, related to translational invariance, are excluded.
For temperatures higher than the Debye temperature ($\theta_{\rm D}=281$\,K for Nb and $261$\,K for Zr \cite{Chen2001}), the free energy becomes proportional to the logarithmic moment of the vibrational DOS, leading to
\begin{equation}
	F_{\rm vib}(T) = (3N-3) k T \ln{ \left( \frac{k T}{\hbar \omega_{\rm log}} \right) },
	\label{eq:Fvib_HT}
\end{equation}
where the logarithmic average of the phonon frequencies $\omega_{\rm log}$ is defined as
\begin{equation}
	\ln{\left( \omega_{\rm log} \right)} =  \frac{1}{3N-3} \sum^{3N-3}_{\vec{q},s}
		\ln{\left( \omega_{\vec{q}}^s \right)}.
	 \label{eq:eq_16}
\end{equation}
This high temperature approximation leads to excess free energies varying linearly with temperature
as the $\ln{(kT)}$ term entering Eq. \ref{eq:Fvib_HT} cancels out in the weighted difference
defining the excess quantity.

In this work, the DFPT method is employed to obtain the harmonic vibrational spectra as described in~\S\ref{subsec:vib}. As shown in \ref{sec:annex2}, the phonon dispersions obtained in pure Nb and Zr
are in good agreement with experiments, as well as with \abinit calculations relying on other modeling approaches. Using single impurity calculations
with bcc ZrNb$_{N-1}$ ($N=12$, 16, 27, and 54) and hcp Zr$_{N-1}$Nb ($N=16$, 36, and 54) supercells, the vibrational excess free energies of the $\beta$ and $\alpha$ solid solutions 
are determined for different concentrations.
The corresponding contribution to the solution free energy is shown in Figures \ref{fig_5}c and d.
This contribution of lattice vibrations is much larger than the one of electronic excitations, 
with both contributions significantly higher in hcp Zr$_{N-1}$Nb than in bcc ZrNb$_{N-1}$ compounds.

Above the Debye temperature, this vibration contribution to the solution free energy
is proportional to temperature (Fig. \ref{fig_5}c and d), 
in agreement with the high temperature approximation \eqref{eq:Fvib_HT}.
Therefore, we interpolate it as a linear function of temperature, 
which is clearly sufficient above 100\,K.
The coefficients $\Omega_{\rm vib}/kT$ thus obtained for each supercell are shown in the insets
of Fig. \ref{fig_5}. 
We deduce from these calculations the vibrational free energies of the $\beta$ and $\alpha$
solid solutions, assuming a linear variation with the concentration (Tab. \ref{tab:Esol}). 
For the $\beta$ Nb-rich phase, the contributions of lattice vibrations to the solution free energy
is small  and does not show any clear variation with the concentration (Fig. \ref{fig_5}c). 
We therefore take it constant: $\Omega^{\beta}_{\rm vib}=-0.41\,kT$. 
For the $\alpha$ Zr-rich phase, this contribution is also negative, but with a much larger absolute value 
(Fig. \ref{fig_5}d). 
In the limit of infinite dilution, we obtain $\Omega^{\alpha}_{\rm vib}(x_{\rm Nb}\to0)=-4.59\,kT$

\section{Solubility limits \label{sec:sol_lim_2}}

\begin{figure}[!bt]
	\centering
	(a) \includegraphics[width=0.9\linewidth]{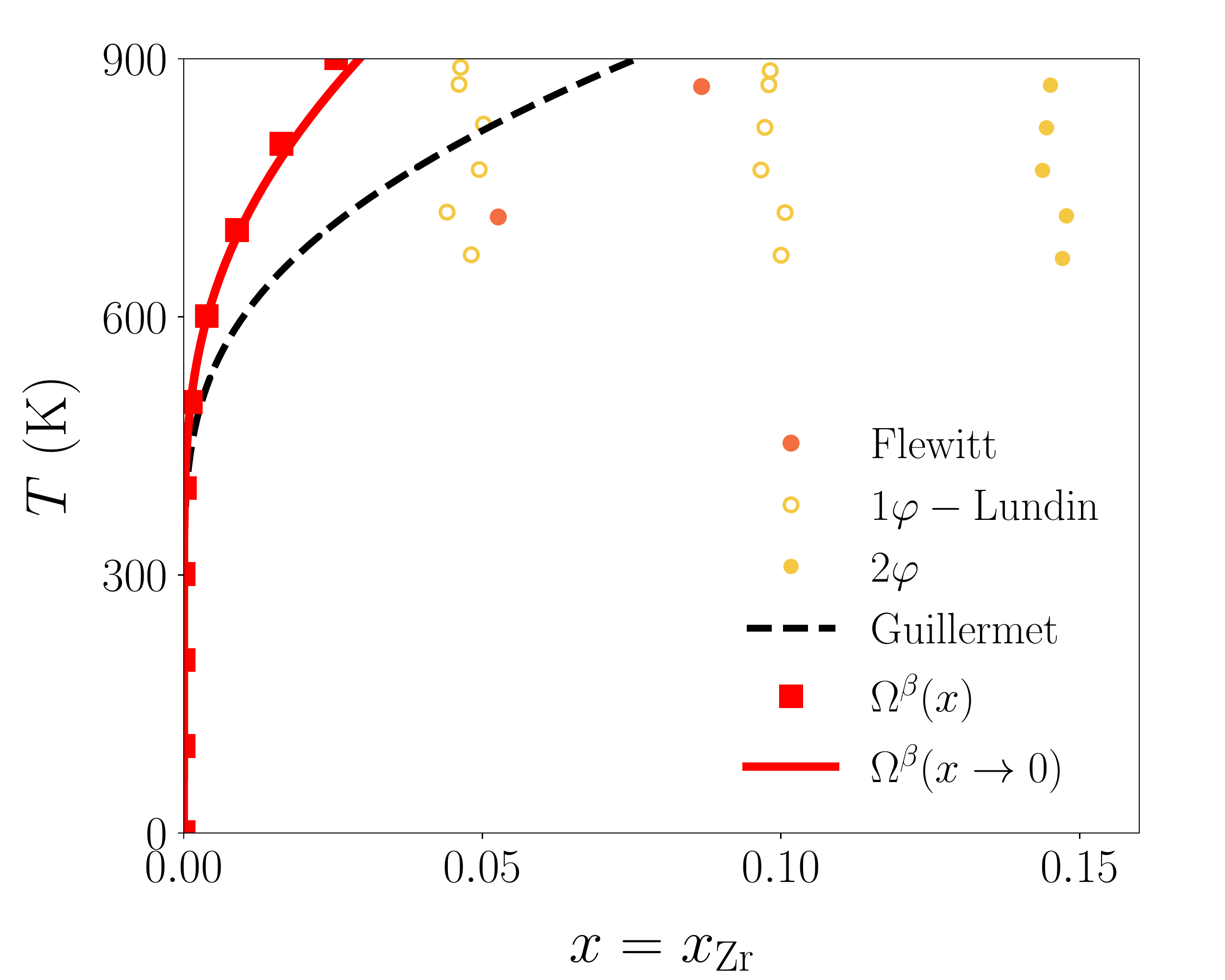} 
	(b) \includegraphics[width=0.9\linewidth]{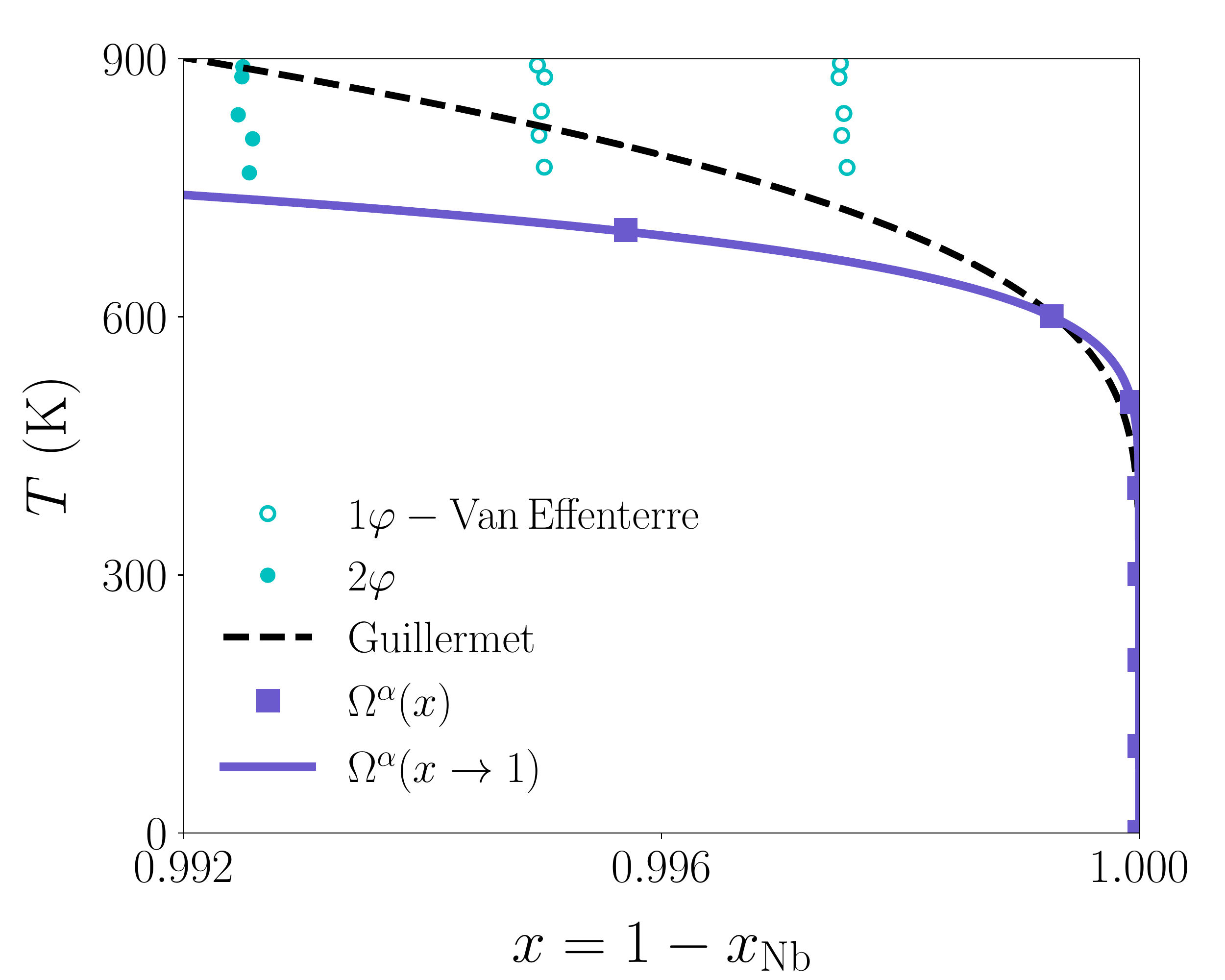}
	\caption{Solubility limits 
	(a) $x^{\beta}$ in the bcc Nb-rich and (b) $x^{\alpha}$ in the hcp Zr-rich solid solutions
	with solution free energies depending on the temperature (Tab. \ref{tab:Esol}).
	The solubility limits obtained by considering all contributions to the solution free energies, 
	as well as their concentration dependence, are shown with colored squares.
	The solid lines correspond to the dilute limit approximation (Eq. \ref{eq:solubility_mix_dilute})
	without the electronic contribution.
	The obtained solubility limits are compared to the Calphad assessment of Guillermet \cite{Guillermet1991}
	and to the experimental limits determined by Flewitt \cite{Flewitt1972}.
	The one phase ($1\phi$) and two phases ($2\phi$) determined experimentally by Lundin \cite{Lundin1960}
	and by Van Effenterre \cite{VanEffenterre1972} are also represented.
	}
	\label{fig:solubility_T} 
\end{figure}

With these electronic excitations and lattice vibrations contributions deduced from \abinit calculations (Tab. \ref{tab:Esol}),
the solution free energies of the $\beta$ and $\alpha$ solid solutions are now temperature-dependent. 
We can use the thermodynamic model described in section \ref{sec:thermo_model}
to obtain the corresponding solubility limits. 
Results are shown in Fig. \ref{fig:solubility_T}, with colored squares when all contributions
with their concentration dependence are considered. 
As the contribution of electronic excitations is much smaller than the other contributions
(Fig. \ref{fig_5}), it can be safely neglected without influence on the resulting solubility limits. 
The concentration dependence of the solution energy can also be ignored, 
using the dilute approximation given by Eq. \eqref{eq:solubility_mix_dilute}. 
This does not impact the solubility limit in the $\alpha$ Zr-rich phase, 
whereas this leads to only a slight variation of the solubility limit in the $\beta$ Nb-rich phase
for the highest temperatures close to 900\,K. 
With these approximations, the solution free energies of the two phases are simply given by
\begin{align*}
	\Omega^{\beta }(T) &= 0.312\,\textrm{eV} - 0.41 \, kT, \\
	\Omega^{\alpha}(T) &= 0.632\,\textrm{eV} - 4.59 \,  kT.
\end{align*}

As shown in Fig. \ref{fig:solubility_T}, 
the solubility limits agreement with the Calphad assessment of Guillermet \cite{Guillermet1991}
is significantly improved
when the non-configurational entropy contribution is accounted for. 
In the Nb-rich part, the finite temperature excitations are not sufficient to fully recover the Calphad result.
The calculated vibration contribution to the solution free energy, $0.41\,k$,
is too low compared to the one required to recover Calphad result, $1.5\,k$.
On the other hand, in the Zr-rich part, the solubility limit deduced from our \abinit modeling
is higher than the Calphad one because of a larger entropy contribution, $4.59\,k$,
than the one corresponding to the Calphad solubility limit, $3.6\,k$. 
Therefore, only a semi-quantitative agreement could be obtained between our \abinit modeling 
and the Calphad assessment of Guillermet \cite{Guillermet1991} and Lafaye \cite{Lafaye2017}.
Both approaches lead to very close solution energies in both $\alpha$ and $\beta$ phases,
but the same agreement could not be obtained on the entropy part. 
As shown by our \abinit calculations, this entropy contribution is mainly due to lattice vibrations
and is significantly higher in the $\alpha$ than in the $\beta$ solid solution.
The comparison with the rough experimental data \cite{Flewitt1972,Lundin1960,VanEffenterre1972} 
leads to the same semi-quantitative agreement: our model slightly underestimates the solubility limit
in the $\beta$ phase (Fig. \ref{fig:solubility_T}a) 
and overestimates it in the $\alpha$ phase (Fig. \ref{fig:solubility_T}b).
Obviously, the Calphad model of Guillermet \cite{Guillermet1991} 
which has incorporated this information in his assessment is closer to experimental data.

One could wonder if a better agreement could be obtained between the \abinit modeling 
and the Calphad assessment, and thus the experimental data. 
In the $\beta$ phase, the vibrational part $\Omega^{\beta}_{\rvib}$ given by the \abinit calculations is small. 
It has been obtained by fitting vibrational free energies calculated with DFPT for 4 different supercells
(Fig. \ref{fig_5}c).  The values corresponding to these 4 impurity calculations show some scattering, 
with the largest contribution, $0.86\,kT$, obtained for the bcc ZrNb$_{15}$ compound, 
and the smallest one, $-0.24\,kT$, for the bcc ZrNb$_{26}$ compound.
The value corresponding to the Calphad limit, $1.5\,k$, is therefore not far from the confidence range of the \abinit calculations. 
In the $\alpha$ phase, $\Omega^{\alpha}_{\rvib}$ 
has been obtained by considering only the two hcp Zr$_{35}$Nb and hcp Zr$_{53}$Nb supercells.
The results obtained for the hcp Zr$_{15}$Nb structure has been excluded from the fit (Fig. \ref{fig_5}d),
as it corresponds to a higher concentration than the solubility limit.
If we were to consider it, a lower value of $\Omega^{\alpha}_{\rvib}$ would have been obtained, 
leading to a better agreement with the value corresponding to the Calphad solubility limit. 
Ideally, one would rather fit $\Omega^{\alpha}_{\rvib}$ on supercells corresponding to lower concentrations. 
However, phonon calculations of the vibrational excess free energy for supercells with more than $N=54$ lattice sites
are too computationally demanding using the DFPT method, and thus appear out of reach. 
Finally, anharmonicity, which is not accounted for in our thermodynamic model, may also explain 
some of the discrepancy. As a first step, volume expansion effects could be accounted for using the quasi-harmonic approximation, although anharmonic contributions have usually only a marginal contribution 
to the phase diagram \cite{Ozolins2001}.

\section{Conclusion}

The thermodynamic properties of the Zr-Nb alloy below its monotectoid temperature ($T\sim890$\,K)
have been determined from \abinit calculations
considering both hcp Zr-rich and bcc Nb-rich solid solutions.
Density functional theory calculations performed for various configurations representative 
of these two phases show that they are simple unmixing systems.
Incorporating the obtained solution energies in a mean field thermodynamic model, 
we have obtained solubility limits which are much lower than the experimental ones. 
To improve the agreement, one needs to include non-configurational entropy 
when defining the solution free energies of the $\alpha$ and $\beta$ phases.
Whereas the electronic free energy can be safely ignored, 
atomic vibrations lead to an important contribution as we have determined within the harmonic approximation 
using DFPT calculations.
Thus, we have obtained solution free energy which depends linearly on the temperature. 
The corresponding solubility limits are in much better agreement with experimental data.
Clearly, atomic vibrations need to be accounted for to obtain a quantitative description 
of the Zr-Nb thermodynamics.

\appendix

\section{Electronic excitations calculations \label{sec:annex1}}

Electronic free energy (Eq. \ref{eq:Fel}) of pure bcc Nb has been seen to be sensitive
to the choice of the supercell and the $\vec{k}$-point mesh.
Different results were initially obtained between calculations using a supercell 
built by homothetic transformations of the conventional cubic cell ($N=2$, 16 and 54 atoms)
or a supercell based on the rhombohedral primitive lattice ($N=1$, 8 and 27 atoms).
The same values could finally be obtained for pure Nb with these two supercells,
using a special $\vec{k}$-point mesh for the rhombohedral cell.
This special mesh is made of two superimposed grids shifted by half a grid step in each direction
and is equivalent to the $\vec{k}$-point mesh used for cubic cells.
On the other hand, the cohesive energy has not shown such a sensitivity to the supercell and the $\vec{k}$-point mesh.

\section{Phonon calculations \label{sec:annex2}}
  
\begin{figure}[!bth]\centering
	\subfloat[bcc Nb]{\includegraphics[width=\linewidth]{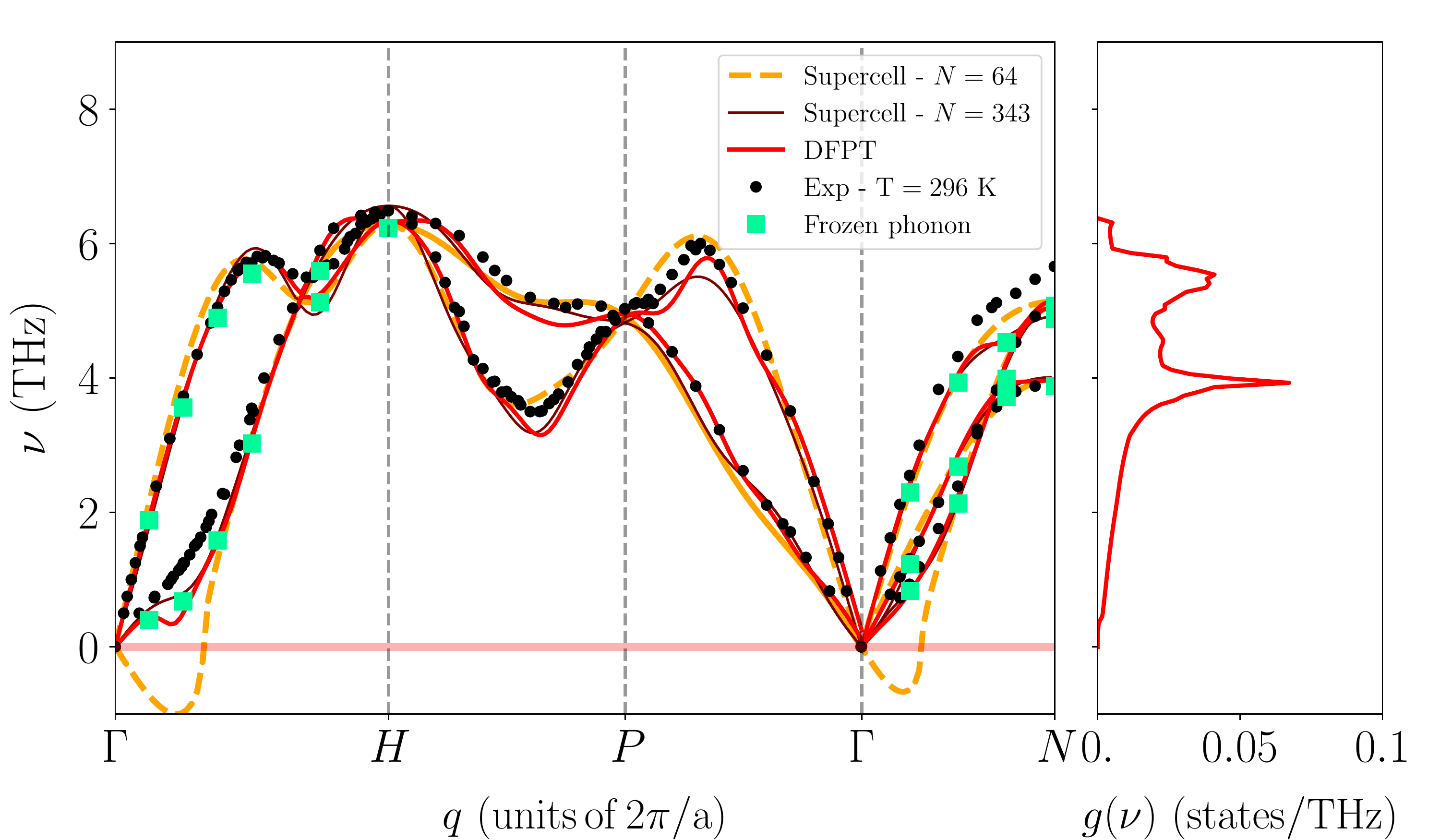}}\\
	\subfloat[hcp Zr]{\includegraphics[width=\linewidth]{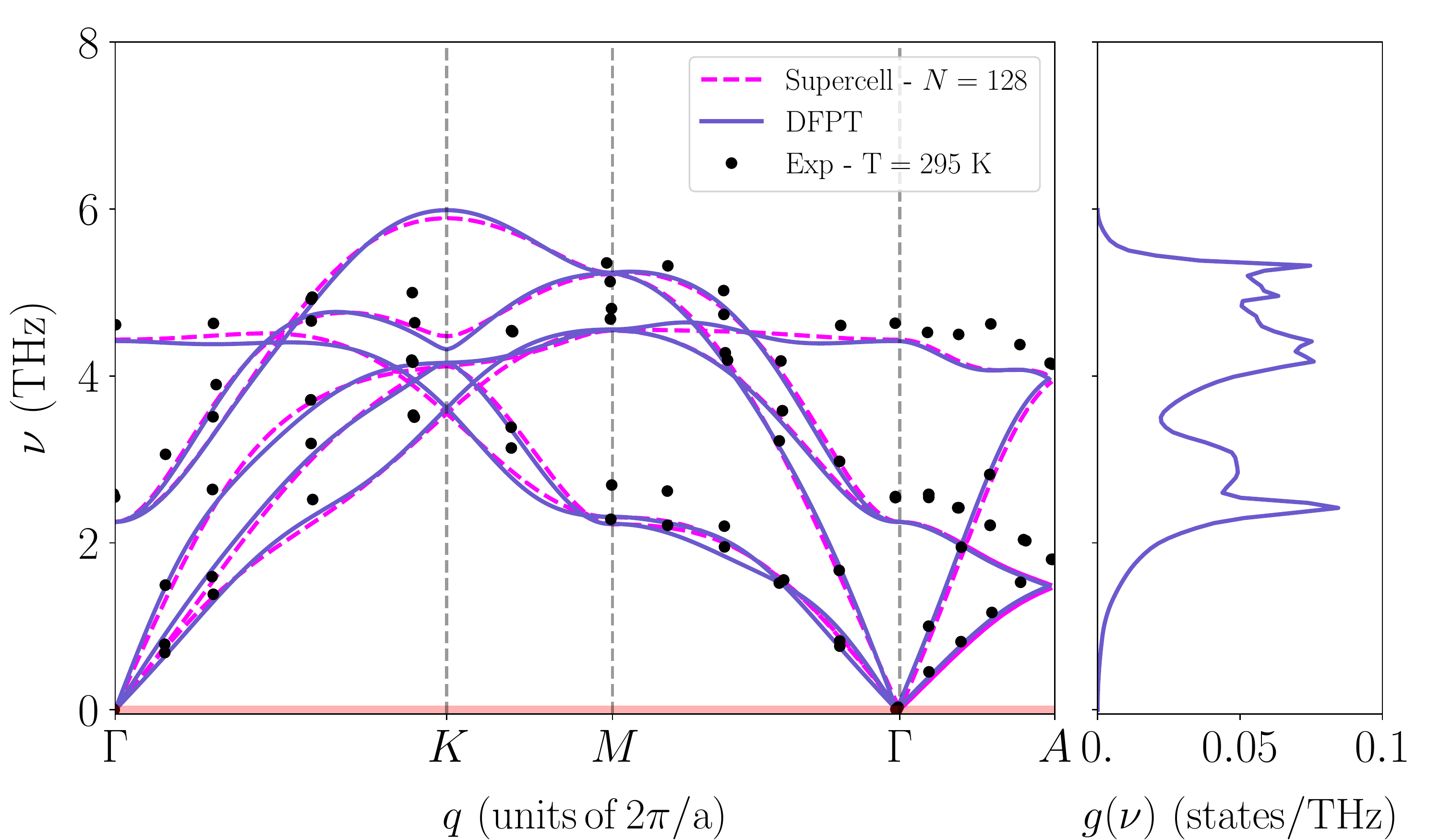}}
	\caption{Phonon dispersion spectra for (a) bcc Nb and (b) hcp Zr.
	The full spectrum (lines) are calculated at $T=0$\,K 
	using either finite differences with supercells of different sizes ($N$ is the number of atoms)
	or the DFPT approach.
	Frozen phonon calculations along high symmetry directions are shown by green squares for bcc Nb.
	The experimental spectra \cite{Powell1977, Stassis1978} measured at $T=296$\,K are shown with black dots. 
	The theoretical phonon DOS (DFPT) are represented on the right side. }
	\label{fig:phonon} 
\end{figure}

The phonon dispersion spectra necessary to obtain the vibrational free energy are calculated with 
the density functional perturbation theory (DFPT) \cite{deGironcoli1995}.
A $12\times12\times12$ mesh of $\vec{q}$ wave vectors is used for the bcc Nb elementary cell.
Such a dense grid is necessary because of the long range behavior of the force constants 
which extend up to the 13$^{\rm th}$ nearest neighbors \cite{Nakagawa1963,Powell1968}. 
We check the convergence with respect to this grid density by comparing 
DFPT results with frozen phonon calculations \cite{Baroni2001} for $\vec{q}$ points lying on high symmetry directions. 
Supercells commensurate with the wave vector $\vec{q}$ are used in such calculations
and the phonon frequencies are deduced from the energy variation when a displacement 
corresponding to the phonon plane wave is imposed on all atoms in the supercell. 
The perfect agreement observed along $\Gamma H$ and $\Gamma N$ directions between DFPT and frozen phonon calculations
(Fig. \ref{fig:phonon}a) ensures that the $\vec{q}$ points grid of DFPT calculations is dense enough.

Instead of DFPT, one can use a large enough supercell and calculate all force constants by finite differences 
\cite{Togo2015}. 
The Nb phonon dispersion spectrum thus obtained with a $4\times4\times4$ bcc supercell (64 atoms)
exhibits negative eigenvalues which correspond to imaginary frequencies (Fig. \ref{fig:phonon}a).
These negative phonon branches arise from the supercell being too small compared to the interaction range
of the force constants. 
Using a larger $7\times7\times7$ bcc supercell (343 atoms), the negative branches disappear 
and the obtained phonon spectrum agrees with the DFPT one.
As the interaction range of the force constants is smaller in hcp Zr, 
well converged phonon spectra are obtained with smaller supercell.
A $4\times4\times4$ hcp supercell containing 128 atoms is enough 
to recover the phonon spectrum obtained with DFPT using a  $6\times6\times5$ $\vec{q}$ mesh
(Fig. \ref{fig:phonon}b).

The $7\times7\times7$ bcc supercell necessary for calculations in the Nb-rich $\beta$ phase
leads to prohibitive CPU cost 
when a solute atom is inserted in the simulation box. 
Therefore, the DFPT method has been preferred to obtain 
the variations of the vibrational entropy with solute concentration.

We can compare the obtained phonon spectra with the experimental data \cite{Powell1977,Stassis1978},
although experiments have been performed at room temperature. 
The agreement is excellent for Nb (Fig. \ref{fig:phonon}a). 
On the other hand, a difference exists in Zr for the highest frequency at $\vec{q}=K$  (Fig. \ref{fig:phonon}a).
The same discrepancy between the theoretical and the experimental spectra was observed in \cite{Hao2008,Hu2011}, 
using the finite difference method and a smaller $3\times3\times3$ hcp supercell with \vasp code.
Except for this wave vector $K$, the agreement with experimental frequencies is good.

\vspace{0.5cm}
\linespread{1}
\small

\textbf{Acknowledgments} -
This work was performed using HPC resources from GENCI-CINES and -TGCC (Grant 2016-096847).
This work is funded by the project ``Transport et Entreposage'' of the French nuclear institute.
The authors are grateful to Fabien Bruneval for his help with the electronic structure calculations.
They also would like to thank Bernard Legrand, Francois Willaime and Émile Maras for fruitful discussions.
\section*{References}

\bibliographystyle{elsarticle-num}
\bibliography{cottura2017}
\end{document}